\documentclass[a4paper,11pt]{article}
\pdfoutput=1
\usepackage{verbatim}
\usepackage{jcappub}
\usepackage[T1]{fontenc} 
\usepackage[utf8]{inputenc}
\usepackage[T1]{fontenc} 
\usepackage{natbib, hyperref}
\usepackage[varg]{txfonts}
\usepackage{caption}
\usepackage{subcaption}
\usepackage{comment}
\usepackage{blindtext}
\usepackage{autobreak}
\usepackage{tikz}
\usetikzlibrary{matrix}
\usepackage{bm}
\usepackage{amsmath}
\usepackage{float}
\usepackage{color}
\usepackage{xcolor}
\usepackage{appendix}
\usepackage{hyperref}
\usepackage{multirow}  
\usepackage{physics}
\usepackage{cancel}
\usepackage[varg]{txfonts}
\usepackage{graphicx}

\allowdisplaybreaks

\newcommand{\Hbb}{\mathbb{H}}
\newcommand{\Qbb}{\mathbb{Q}}
\newcommand{\Rbb}{\mathbb{R}}
\newcommand{\Jbb}{\mathbb{J}}
\newcommand{\Sbb}{\mathbb{S}}

\title{Covariant cosmography in the presence of local structures: comparing exact solutions and perturbation theory}

\author[]{Maharshi Sarma$^1$, Christian Marinoni$^1$, Basheer Kalbouneh$^1$, Chris Clarkson$^{2,3}$ and Roy Maartens$^{3,4}$}   

\affiliation[1]{\small{Aix Marseille Univ, Universit\'e de Toulon, CNRS, CPT, Marseille, France}}
\affiliation[2]{\small{Department of Physics \& Astronomy, Queen Mary University of London, London E1 4NS, United Kingdom}}
\affiliation[3]{\small{Department of Physics \& Astronomy, University of the Western Cape, Cape Town 7535, South Africa}}
\affiliation[4]{\small{National Institute for Theoretical \& Computational Sciences, Cape Town 7535, South Africa}}

\abstract{Recent observational evidence of axially symmetric anisotropies in the local cosmic expansion rate motivates an investigation of whether they can be accounted for within the Lemaître– Tolman-Bondi (LTB) framework with an off-center observer. Within this setting, we compute the exact relativistic luminosity distance via the Sachs equation and compare it with the approximate expression  obtained from the covariant cosmographic approach (including Hubble, deceleration, jerk and curvature parameters).  
This comparison allows us to  identify the regimes in which the covariant cosmographic method remains reliable. 

In addition, we compare the LTB relativistic distance for small inhomogeneities with the corresponding result derived from linear perturbation theory (LPT) in the standard cosmological model. This analysis establishes a precise correspondence between the LTB and LPT approaches, offering a consistent dictionary for the interpretation of the observed anisotropies of the large-scale gravitational field. 

We test luminosity distance reconstructions in a spherically symmetric overdensity with an off-center observer. For moderate  central density contrasts ($\delta_c\lesssim1$), LPT reproduces the exact distance within $10\%$ for observers inside the typical size of the structure. However, Covariant Cosmography (CC) extends this regime of validity upto $\delta_c\lesssim 2.5$. At larger radii, the situation reverses: for observers at three times the characteristic size, LPT remains accurate up to $\delta_c\lesssim 3$, while CC already exceeds $10\%$ error for $\delta_c\gtrsim 1.5$. At sufficiently large distances from the structure, both methods converge to the exact solution. Thus, CC is essential for accurate distance estimates near dense regions, while LPT remains reliable at larger separations. 

This analysis will be instrumental in interpreting expansion-rate anisotropies, facilitating investigations of the local Universe beyond the FLRW framework with a fully non-perturbative metric approach.}

\begin{document}
\maketitle
\flushbottom

\section{Introduction}

Cosmology currently contends with perplexing anomalies that have proven resistant to resolution within the framework of the standard model
\cite{DiValentino:2021izs, Perivolaropoulos:2021jda, Schoneberg:2021qvd, Abdalla:2022yfr, Capozziello:2024stm, Bengaly:2024ree}. 
Among these anomalies, the tension between local measurements of the Hubble constant ($H_0$) and its value inferred from the Cosmic Microwave Background (CMB) highlights the need to revisit fundamental assumptions about spacetime geometry---most notably, the cosmological principle, which posits that the universe is homogeneous and isotropic on large scales \cite{Schwarz:2007wf, Kashlinsky:2008ut, Antoniou:2010gw, Cai:2011xs, Marinoni:2012ba, Kalus:2012zu, Wang:2014vqa, Yoon:2014daa, Tiwari:2015tba, Javanmardi2015, Bengaly:2015nwa, Colin:2017juj, Rameez:2017euv, Migkas:2020fza, Migkas:2021zdo, Secrest:2020has, Siewert:2020krp, Luongo:2021nqh, Krishnan:2021jmh, Sorrenti:2022zat, Aluri:2022hzs, Cowell:2022ehf, Hu:2023eyf,Hu:2023jqc, Dainotti:2021pqg,CosmoVerseNetwork:2025alb,Mazurenko:2024gwj,Sah:2024csa,Hu:2024big,Franco:2024dvc,Adam:2024kgs,Lopes:2024vfz,Rameez:2024xsn,Celerier:2024dvs}.

In a series of works \cite{kalbouneh_marinoni_bel_2023, Maartens:2023tib, Kalbouneh:2024, Kalbouneh:2024yjj}, we have investigated the possibility of describing anisotropies in the local cosmic expansion rate within a fully relativistic framework, without resorting to linear perturbation theory approximations---such as small overdensity contrasts or peculiar velocities---and in a model-independent manner, free from \emph{a priori} symmetry constraints on the underlying gravitational field.
The  goal is to characterize the expansion rate on local cosmic scales ($z \lesssim 0.1$ or $r\lesssim 300$ $h^{-1}$ Mpc) more meaningfully and comprehensively than using the $H_0$ parameter of the Standard Model alone. This program involves two critical steps. First, we need to identify and classify deviations from isotropy in the observed redshift-distance relation in an unbiased and model-independent way.  Then, we need to relate these angular distortions to kinematic and dynamic quantities, defined covariantly in arbitrary spacetimes, to gain insight into the cosmic geometry in the observer's surroundings.

The first challenge was tackled by introducing the  expansion rate fluctuation field $\eta$ \cite{kalbouneh_marinoni_bel_2023, Kalbouneh:2024}.   This observable is only sensitive to the angular structure of the fluctuating component in the redshift vs. luminosity distance relation and its possible evolution with redshift. It is only in the presence of angular anisotropies of the luminosity distance, at a given redshift, that $\eta$ deviates from zero. 
As a scalar, Gaussian indicator,  $\eta$ presents several statistical advantages and it lends itself quite readily to a decomposition on a spherical harmonic basis, facilitating straightforward signal interpretation. The additional advantage is that, being a local observable, independent from the assumption of a uniform cosmic background,   any potential systematic errors in measurements of the luminosity distances as a function of redshift, such as Malmquist bias, do not affect the signal of $\eta$.

The second challenge lies in generalizing the concept of the cosmic expansion rate to account for what "expansion" means for an arbitrary observer in a general spacetime. This idea dates back to the seminal work \cite{kristian_sachs_1966}  where possibly non-standard (i.e. anisotropic) expansion rate histories were explored through the tensorial expansion of the luminosity distance with respect to redshift. This approach, refined and extended in subsequent studies  \cite{MacCallum_Ellis_1970,ellis_2009,ellis_1983,ellis85, Hasse:1999,Clarkson_theses_2000,clarkson_maartens_2010,Umeh:2013,heinesen_2021, Maartens:2023tib, Kalbouneh:2024yjj} yields the covariant cosmographic (CC)  parameters,
$\mathbb{H}(\textbf{n})$ (Hubble), $\mathbb{Q}(\textbf{n})$ (deceleration), $\mathbb{J}(\textbf{n})$ (jerk),  $\mathbb{R}(\textbf{n})$ (curvature) and $\mathbb{S}(\textbf{n})$ (snap). 
These parameters depend on the line-of-sight direction $(\textbf{n})$ and characterize the geometry around the observer. At the same time, they are related to the expansion coefficients of the luminosity distance as a function of redshift, which renders them observable. They therefore provide a fully model-independent framework for interpreting cosmic expansion around an observer, without requiring any specific metric or dynamical theory of gravity for their estimation.
Although they are in principle functional degrees of freedom, they have  a finite number of spherical harmonic components. Consequently, the CC  parameters can be estimated  by fitting a limited subset of their multipoles to the data. 
This significant operational advantage is counterbalanced by the fact that modeling the distance–redshift relation over a broad redshift range within covariant cosmography requires an effective description of the cosmic matter field as a coarse-grained, sufficiently smooth, pressure-free fluid (“dust”). As a result, covariant cosmography provides an idealized  description that can be attained only by analyzing data on sufficiently large scales, after neglecting sparse and noisy local data in the observer’s surroundings, as argued by \cite{paper5}.
This effective approach  has been shown to provide unbiased results, as demonstrated both analytically \cite{Kalbouneh:2024} and through simulations  \cite{Adamek:2024hme}.

In this paper, we build on these preliminary analyses by estimating the amplitudes of the covariant cosmographic parameters—including the Hubble, deceleration, curvature, and jerk parameters as determined by an off-center observer in a spherically symmetric Lemaître-Tolman-Bondi (LTB) spacetime \cite{tolman, bondi}.
Interest in this cosmological scenario is motivated by emerging evidence for an approximately axisymmetric structure in the angular anisotropies of the local cosmic expansion rate. A spherical-harmonic decomposition of the expansion-rate fluctuation field reveals a coherent alignment across multipoles: the dipole direction coincides with the quadrupole maximum, the octupole likewise peaks along this same axis,  and both   exhibit approximate axial symmetry about the dipole. 
\cite{kalbouneh_marinoni_bel_2023} identified this pattern at redshifts $z<0.05$ using \textit{Cosmicflows-3} data \cite{Tully_2016}.  Subsequent analyses by \cite{paper5},  based on \textit{Cosmicflows-4} data \cite{Tully:2022rbj},  confirmed that this alignment persists to at least $z<0.1$.
 The simplest scenario in which an axially symmetric anisotropic expansion rate could, in principle, be observed occurs when we, as observers, are positioned at a distance from a massive structure that dominates the kinematics of the local universe. The objective is phenomenological: to develop an analytical framework for interpreting observational estimates of the covariant cosmographic parameters and to assess the viability of this model for the local metric.

The second goal, more theoretical, is to compare the fully relativistic, nonperturbative predictions for the amplitude of the covariant cosmographic parameters, derived within the LTB metric framework, with the approximate results obtained in standard cosmology using linear perturbation theory applied to a spherically symmetric mass in a smoothly expanding, flat Friedman-Robertson-Walker (FRW) background. The objective is to identify subtle relativistic effects occurring on horizon scales that may be overlooked in model-dependent analyses based on the cosmological principle, as well as to uncover the imprints of non-linearities that could bias results obtained using first-order perturbative methods in standard cosmology. 
This comparison is crucial, as the latter paradigm is often employed to identify the region of parameter space in which covariant cosmography remains reliable (see for example  \cite{Kalbouneh:2024, Koksbang_2025,Modan_2024}). It is therefore essential to assess whether, and under what conditions, linear perturbation theory (LPT) provides an unbiased representation of the true fully relativistic LTB metric.

The paper is organized as follows.
    In Section 2, we review the key properties of the LTB cosmology. 
    After introducing the formalism, we detail the calculation of the luminosity distance for an off-center observer within the LTB metric.
    In Section 3, we expand the luminosity distance as a function of redshift upto $\mathcal{O}(z^3)$  and provide analytical expressions for all relevant multipoles of the CC parameters for an off-center LTB observer.
    Section 4 specifies these general expressions by evaluating them for a particular spherically symmetric mass distribution in the universe. The goal is to calculate the accuracy of the CC formalism in reproducing the fully relativistic luminosity distance.
    In Section~5 we construct a correspondence between the relevant parameters of the linearized LTB metric and those of linear perturbation theory in standard cosmology. This correspondence enables us to relate the linearized expressions of the covariant cosmographic parameters in the LTB framework to their counterparts expected in a linearly perturbed FRW background.
    Finally, Section 6 presents our conclusions and outlines directions for future research.

Hereafter, we adopt the Einstein summation convention for repeated indices, in Greek letters (from 0 to 3) and Latin letters (from 1 to 3). We use natural units $(c = 1)$ unless stated otherwise and the metric signature is $(- + ++)$.
 We assume a spherical coordinate system $(t,\chi,\theta,\phi)$. The prime indicates a partial derivative with respect to the radial coordinate $\chi$ and the overdot indicates a partial derivative with respect to the time coordinate  $t$.

\section{LTB cosmology}
The Lema\^itre–Tolman–Bondi metric describes a spherically symmetric  universe with radial inhomogeneities that can be charted using 
spherical spatial coordinates $x^i \equiv (\chi, \theta, \phi) $  that comove  with matter ($\tau$, the proper time of an observer comoving with the matter particle,  coincides with the coordinate time, $x^0 \equiv t$.)
The distance between two infinitesimally close spacetime events  is given by the following line element \cite{bondi}
\begin{equation}
 \dd s^2=-\dd t^2+\alpha^2(t, \chi)\, \dd\chi^2 +A^2(t, \chi)\,(\dd\theta^2+\sin^2{\theta}\, \dd\phi^2).
\label{dsltb}
 \end{equation}
Note that the radial coordinate is not uniquely  determined: the form of the line element remains unchanged under a reparametrization $\chi \rightarrow \chi' = g(\chi)$. Such a coordinate transformation merely rescales the scale factors. 
To eliminate this coordinate freedom, we fix a constant time $t_0$— chosen as the present age of the universe as measured by the off-center observer at $\chi_o$ and arbitrarily specify the function $A(\chi, t_0)$ by setting  $A_0(\chi) \equiv A(\chi, t_0) = \chi$.

To derive the dynamics of the scale factors $\alpha$ and $A$, we further assume that the matter distribution can be modeled as a perfect fluid, with a stress-energy tensor given by $T_m^{\mu\nu} =\rho_m(t,\chi)u^\mu u^\nu$ where the four-velocity is $u^\mu \equiv \dd x^{\mu}/\dd\tau=\delta^\mu_0$. The  transverse scale factor  $A(t,\chi)$ is determined via the Friedmann-like  equation (which can be derived from the $(0,0)-$component of the Einstein field equations \cite{Codur:2021wrt}) 
\begin{equation}
\left( \frac{\dot{A}} {A }\right)^2+\frac{k}{A^2}=\frac{8 \pi G}{3}  \left( \tilde \rho_m+\rho_{\Lambda}\right),
\label{ltb1}
\end{equation}
where $\rho_{\Lambda}$ is the non-dilutive energy density associated with the cosmological constant $\Lambda$ and $\tilde \rho_m$ is the \textit{flat average density} of matter defined by
\begin{equation} 
\tilde \rho_m(t,  \chi) \equiv 3\; \frac{ \int_{0}^{\chi} \rho_m(t,\chi)  A^2(t,\chi) A'(t,\chi) \dd\chi} {A^3(t,\chi)},
\label{md}
\end{equation}
with $\rho_m$ being the physical matter density in the comoving frame. 
The longitudinal scale factor is determined from the $(0,1)-$component of the Einstein field equations \cite{Codur:2021wrt}
\begin{equation}
    \begin{split}
        \alpha(t,\chi)&=\frac{A'(t,\chi)}{\sqrt{1-k(\chi)}}.
    \end{split}
\end{equation}
Here, $k(\chi) <1$ is an arbitrary function of the radial coordinate $\chi$ alone. 

Related to the scale factors  are two useful kinematical quantities, \textit{viz.,} the longitudinal and transverse expansion rates defined as 
\begin{equation}
H_{\|}(t,\chi)\equiv \frac{\dot{\alpha}(t,\chi)}{\alpha (t,\chi)},
\end{equation}
and 
\begin{equation}
H(t,\chi) \equiv \frac{\dot{A}(t,\chi)}{A(t,\chi)}, 
\end{equation}
respectively.
They are  related as 
\begin{equation}
H_{\|}=H+\frac{A}{A'}H'=H+\frac{A^2}{2\dot{A}A'}\left[\frac{8\pi G}{3}\tilde \rho_m'- \left( \frac{k}{A^2} \right)'\right],
\end{equation}
and, at the center of symmetry ($\chi=0$),  the two expressions coincide:  $H_{\|}(0, t)=H(0, t)$.   They are also identical at all times $t$ in the limiting case of an FLRW universe.

By means of the variable $\tilde{\rho}_m$, equation (\ref{ltb1}) can be recast in a form formally analogous to the Hubble expansion rate in the standard cosmological model:
\begin{equation} 
H^2(\chi,t)=H_0^2(\chi)\left[\tilde\Omega_{m0}(\chi)\left(\frac{\chi}{A(t,\chi)} \right)^3 + \Omega_{k0}(\chi)\left(\frac{\chi}{A(t,\chi)} \right)^2 + \Omega_{\Lambda 0}(\chi) \right],
\label{thexp}
\end{equation}
where 
\begin{equation}
    \Omega_{\Lambda 0}(\chi) = \frac{8\pi G}{3H_0^2(\chi)}\rho_\Lambda,
\end{equation}
\begin{equation}
    \Omega_{k0} (\chi) = -\frac{k (\chi)}{H_0^2(\chi) \chi^2},
\end{equation}
\begin{equation}
    \tilde\Omega_{m0} (\chi) = \frac{8\pi G}{3H_0^2(\chi)} \tilde{\rho}_{m0} (\chi),
    \label{oh}
\end{equation}
and satisfy the constraint $\tilde\Omega_{m}+\Omega_{\Lambda}+\Omega_k =1.$ A suffix $0$ means that these quantities are evaluated at present time $t_0$. (\ref{thexp}) has two degrees of freedom which are the two functions $\tilde \rho_{m0}(\chi)$ and $H_0(\chi)$.

In the following, we set  $\Lambda=0$  since our primary interest lies in exploring the structure of the covariant cosmographic functions in LTB spacetime ( see \S \ref{sec 3}), rather than focusing on effective applications to observational data. In addition, this choice allows us to compare the fully relativistic results with the analytical expressions obtained in the linear approximation of the Einstein–de Sitter cosmological model (see \S \ref{linltb}).

\subsection{The angular diameter distance for off-center LTB observers}
Having established the formalism, we now show how an off-center observer in this spherically symmetric gravitational field estimates the angular diameter distance to distant sources.

Consider  light-rays emitted by distant   sources and converging to the off-center observer at position $\chi_o$ at present time $t_0$. 
The null geodesics through the position of the observer are determined by the set of equations
\begin{align}
\dv{k^{\mu}}{\lambda}+\Gamma^{\mu}_{\nu \rho}k^{\nu} k^{\rho}=0,
\end{align}
where $\mu=0,1,2,3$, $k^{\mu}=\dd x^{\mu}/\dd\lambda$  is the tangent to the  geodesics  and $\lambda \in \mathbb{R}$ is an affine parameter that describes the path of the light ray. 
We choose the affine parameter $\lambda$ so that $\lambda  = 0$ at the observer's position, and impose the normalizing condition $k_{\mu}u^\mu = 1$; the timelike part of the photon wave-vector  is past oriented, i.e. pointing from the source to the observer. With these choices, in the rest frame of the observer, the affine parameter $\lambda$ has a simple physical interpretation: it corresponds to the Euclidean distance to a nearby galaxy.

Following the convention of \cite{Alnes} we position a generic observer on an arbitrary axis ($z$-axis) with respect to which the angle $\theta$ is measured, so that its LTB comoving location is fully specified by the coordinates $\chi=\chi_o$ and $\theta=0$. Owing to the axial symmetry about the $z$-axis the angle $\phi$ becomes redundant and the null geodesic path is entirely determined by only three equations
 \begin{align}
 \dv{k^0}{\lambda}& =-\alpha \dot{\alpha}(k^1)^2-A\dot{A}(k^{2})^2, \label{offko} \\
 \dv{k^1}{\lambda}& =-\frac{\alpha'}{\alpha}(k^{1})^2-2\frac{\dot{\alpha}}{\alpha}k^0k^1+A'\frac{A}{\alpha^2}(k^2)^2, \label{offk1}\\
 \dv{k^2}{\lambda}& =-2\frac{\dot{A}}{A}k^0k^2-2\frac{A'}{A}k^1 k^2. \label{offk2}
 \end{align}
The last equation expresses the conservation of the angular momentum of the photon $J$,
 \begin{align}
 \dv{}{\lambda} \left( A^2k^2\right) =0\equiv \frac{\dd J}{\dd \lambda}.
 \end{align}
The null-geodesic constraint $k^{\mu}k_{\mu}=0$ provides  the additional condition
\begin{align}
 -(k^0)^2+\alpha^2 (k^1)^2+ \left(\frac{J}{A}\right)^2=0,
 \end{align}
from which one obtains that along light rays
\begin{align}
k^1 & = \frac{k^0}{\alpha} \sqrt{1- \left (\frac{J}{Ak^0}\right) ^2},  \label{kk1} \\
k^2 & =\frac{J}{A^2}. \label{kk2}
\end{align}

\begin{figure}
    \centering
    \includegraphics[scale=0.5]{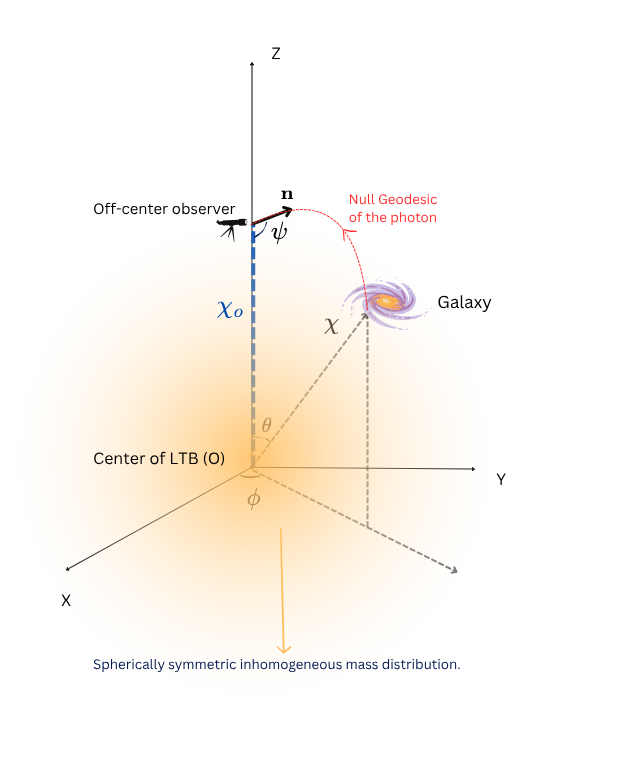}
    \caption{Off-center LTB metric configuration. The observer is located at a distance of $\chi_o$ from the center along the z-axis. The angle $\psi$ denotes the separation between the direction of the center and the observer’s line of sight $\boldsymbol{n}$.}
    \label{fig:2d}
\end{figure}

The constant angular momentum $J$ 
provides a practical way to label the direction of arrival of photons at the position of the off-center observer.

At the position of an off-center observer (with \(\theta_o = 0\)), one has 
\begin{align}
    J &= \chi_o \sin \psi.
\end{align}
The null 4-momentum at the event of observation is 
\begin{align}
    k^\mu(\lambda=0) &= \begin{pmatrix}
        -1\\
        \frac{-1}{\alpha}\sqrt{1-\left(\frac{J}{A} \right)^2}\\
        \frac{J}{A^2}\\
        0
    \end{pmatrix}.
\end{align}

In contrast, for an observer located at the center (i.e., \(\chi_o=0) \), $J=0$. Moreover, light travels radially, implying $J/A^2=0$. 
In Appendix A  we show that the angular diameter distance for the off-center observer can be calculated as 
 \begin{equation}
 d_A(\lambda)=\sqrt{ {\rm det}\,(\mathcal{D}_{IJ}(\lambda))},
 \label{dida}
 \end{equation}
where  $\mathcal{D}^I_{\phantom{I}J}$ is a  $2 \times 2 $ matrix that solves the initial value problem ({\it Sachs equations})
\begin{eqnarray}
\dv[2]{}{\lambda}\mathcal{D}_{IJ}(\lambda)  & = & \tau_{IK} \mathcal{D}^{K}_{\phantom{K}J},  \nonumber \\
\left . \mathcal{D}_{IJ}\right|_0 & = & 0,  \nonumber \\
\left . \dv{}{\lambda}\mathcal{D}_{IJ} \right |_0 & = & \delta_{IJ}.
\label{sachs}
\end{eqnarray}
where $\tau^I_{\phantom{I}J}$ is  the {\it optical tidal tensor}, a  symmetric matrix that connects the evolution of the light bundle with the curvature of spacetime.  We find (see Appendix A)  that  in an LTB cosmology where gravity is sourced by pressureless matter and a cosmological constant, the optical matrix is diagonal and it is given by  
\begin{equation}\label{tidal tensor}
\tau^I_{\phantom{I} J}=-4\pi G \rho_m \left[ \left(1+z\right)^2+ \left ( \frac{J}{A} \right )^2 \left( \frac{\rho_m-\tilde \rho_m}{\rho_m} \right)\left(-1\right)^{I-1}\right] \delta^{I}_{\phantom{I}J}.
\end{equation}

The second term in parentheses is the Weyl focusing term, indicating that the beam is sheared proportionally to $(\rho_m - \tilde{\rho}_m)/\rho_m$. This term does not vanish at the observer's off-center position,  since the local density at  $\chi_o$ differs from the spherical average enclosed within that radius.
There are special cases where shearing is negligible and the beam expands/contracts isotropically. Indeed, the contribution of the Weyl focusing  vanishes in the spherically symmetric configuration, i.e. when the observer sits at the center of the LTB metric $(J=0)$, or if the flat average density $\tilde \rho_m$ equals the local matter density $\rho_m$ as  is the case,  for example, in the standard FLRW metric. 
In these cases it is straightforward to show that the solution of (\ref{sachs})  leads to  the standard formula $d_A(\lambda)=A(t(\lambda), \chi(\lambda)).$

\section{Covariant cosmographic (CC) parameters for an off-center observer in the LTB metric}\label{sec 3}
  Interestingly,  the luminosity distance  $d_L=(1+z)^2d_A$ to a source can be estimated in a fully model-independent way, without requiring any a-priori assumption about the metric of the spacetime. This is achieved  by Taylor expanding the luminosity distance of a light source at a redshift $z$
along the line of sight direction specified by the unit vector $\textbf{n}$ as
\begin{align}\label{dlcc}
        d_L(z,\textbf{n}) &= d_L^{(1)}(\textbf{n})z + d_L^{(2)}(\textbf{n})z^2 + d_L^{(3)}(\textbf{n})z^3 + d_L^{(4)}(\textbf{n})z^4 + \mathcal{O}(z^5).
\end{align}

The expansion coefficients $d^{(i)}_L$  depend on the position of the observer, the time of
observation, and also the line of sight $\textbf{n}$. More importantly, they incorporate information about  the local 
metric in the surroundings of the observer.  In fact, as shown by \cite{kristian_sachs_1966,ellis_2009, MacCallum_Ellis_1970,Clarkson_theses_2000,clarkson_maartens_2010,heinesen_2021, Maartens:2023tib, Kalbouneh:2024yjj}, they can be related to the matter frame {\it covariant cosmographic parameters,} $\mathbb{H}_o(\textbf{n})$ (Hubble), $\mathbb{Q}_o(\textbf{n})$ (deceleration), $\mathbb{J}_o(\textbf{n})$ (jerk), $\mathbb{R}_o(\textbf{n})$ (curvature) and $\mathbb{S}_o(\textbf{n})$ (snap)
\begin{equation}
\begin{cases}
    \begin{split}
        d_L^{(1)}(\textbf{n}) &= \frac{1}{\mathbb{H}_o(\textbf{n})}, \\
        d_L^{(2)}(\textbf{n}) &= \frac{1 - \mathbb{Q}_o(\textbf{n})}{2\mathbb{H}_o(\textbf{n})}, \\
        d_L^{(3)}(\textbf{n}) &= \frac{\mathbb{Q}_o(\textbf{n}) - \mathbb{J}_o(\textbf{n}) + \mathbb{R}_o(\textbf{n}) + 3\mathbb{Q}_o^2(\textbf{n}) - 1}{6\mathbb{H}_o(\textbf{n})}, \\
        d_L^{(4)}(\textbf{n}) &=\frac{1}{24 \mathbb{H}_o(\textbf{n})}\bigg(2-\Qbb_o(\textbf{n})\big(2+15\Qbb_o(\textbf{n})+15\Qbb_o^2(\textbf{n})\big),\\&\hspace{3cm}
        +5\Jbb_o(\textbf{n})(1+2\Qbb_o(\textbf{n})) -2\Rbb_o(\textbf{n})(1+3\Qbb_o(\textbf{n}))+\Sbb_o(\textbf{n})\bigg)
    \end{split}
    \label{eq:cosmographic_parameters}
\end{cases}
\end{equation}

The covariant cosmographic parameters up to the fourth order expansion in redshift are defined by \cite{Maartens:2023tib}
\begin{align}
\mathbb{H} &\circeq k_{\mu}k_{\nu}\Theta^{\mu\nu},\label{cc1} \\
\mathbb{Q} &\circeq -3 + \frac{k_{\mu}k_{\nu}k_{\alpha}\nabla^{\alpha}\Theta^{\mu\nu}}{\mathbb{H}^2}, \label{cc2}\\
\mathbb{R} &\circeq 1 + \mathbb{Q} - \frac{k_{\mu}k_{\nu}R^{\mu\nu}}{2\mathbb{H}^2}, \label{cc3} \\
\mathbb{J} &\circeq -10\mathbb{Q} - 15 + \frac{k_{\mu}k_{\nu}k_{\alpha}k_{\beta}\nabla^{\alpha}\nabla^{\beta}\Theta^{\mu\nu}}{\mathbb{H}^3}, \label{cc4}\\
\Sbb &\circeq 113+17\Jbb+115\Qbb+10\Qbb^2-8\Rbb-\frac{ k_{\mu}k_{\nu}k_{\alpha} \nabla^{\alpha}R^{\mu \nu}}{\Hbb^2}-\frac{k_{\mu}k_{\nu}k_{\alpha}k_{\beta}k_{\gamma}\nabla^{\alpha}\nabla^{\beta}\nabla^{\gamma}\Theta^{\mu\nu}}{\Hbb^4},\label{cc5}
\end{align}
where $\circeq$ indicates that all the quantities are evaluated at the event of observation $o$ and 
$\Theta^{\mu\nu} \equiv \nabla^{\mu}u^{\nu}$ is the matter expansion tensor. Note that in the framework of covariant cosmography, the observable quantity is $\mathbb{H}_o$ and not $H_0$ \cite{Maartens:2023tib}.
These parameters can be reconstructed from observation and provide a glimpse of the local underlying metric. Here we take the opposite approach and we  evaluate them for an off-center observer in the  LTB spacetime.  

To this end, and following \cite{Kalbouneh:2024yjj},  we  define the useful quantities 
\begin{align}
\mathbb{X}^{(1)} & \equiv k_{\mu}k_{\nu}\Theta^{\mu\nu},\label{X1} \\
\mathbb{X}^{(2)} & \equiv k_{\mu}k_{\nu}k_{\alpha}\nabla^{\alpha}\Theta^{\mu\nu}, \label{X2}\\
\mathbb{X}^{(3)} & \equiv k_{\mu}k_{\nu}k_{\alpha}k_{\beta}\nabla^{\alpha}\nabla^{\beta}\Theta^{\mu\nu}, \label{X3}  \\
\mathbb{X}^{(4)} & \equiv k_{\mu}k_{\nu}k_{\alpha}k_{\beta}k_{\gamma}\nabla^{\alpha}\nabla^{\beta}\nabla^{\gamma}\Theta^{\mu\nu}, \label{X4}  \\
\mathbb{Y}^{(1)} & \equiv k_{\mu}k_{\nu} R^{\mu \nu},\\
\mathbb{Y}^{(2)} & \equiv k_{\mu}k_{\nu}k_{\alpha} \nabla^{\alpha}R^{\mu \nu}.
\end{align}
which can be straightforwardly decomposed into spherical harmonics. 
Indeed, unlike standard functional parameterizations that involve
infinite degrees of freedom, the degrees of freedom for each covariant cosmographic parameter are finite and are represented by their multipoles \cite{Hasse:1999,Clarkson_theses_2000,clarkson_maartens_2010,Umeh:2013,heinesen_2021,Maartens:2023tib}. 
A limited set of   multipoles allows us to fully reconstruct the functional form of the Hubble, deceleration, curvature, jerk and snap 
parameters.

Since we  deal with  an axially symmetric gravitational field, the covariant cosmographic parameters that are measured by an off-center LTB observer  will depend only on the angle $\psi$ between the line of sight and the axis of symmetry. If the z-axis is in the direction of the center, it is thus enough to decompose them into multipoles using as basis the  Legendre polynomials $P_\ell$.
As a result,  the spherical harmonic expansion coefficients   $\mathbb{X}^{(i)}_{\ell m}$ 
are zero for $m \neq 0$, and $\mathbb{X}^{(i)}_{\ell 0}=[4\pi/(2\ell+1)]^{1/2} \mathbb{X}^{(i)}_\ell$, where 	
\begin{equation}
\mathbb{X}^{(i)}_{\ell}(z)=\frac{2\ell+1}{2} \int_{-1}^{1} \mathbb{X}^{(i)} \; P_\ell(\cos\psi)\;\mathrm{d}(\cos\psi)\,.
\end{equation}

The relevant (non-zero) multipoles are
\begin{align}
   \mathbb{X}^{(1)}_{0}\circeq & \,\frac{1}{3} (2 H+H_{\|}),\\
   \mathbb{X}^{(1)}_{2}\circeq & \,-\frac{2}{3} (H-H_{\|}),\\
   \mathbb{X}^{(2)}_{0}\circeq & \,\frac{1}{3} \left(4 H^2-2 \dot{H}+2 H_{\|}^2-\dot H_{\|}\right),\\
\mathbb{X}^{(2)}_{1} 
                         \circeq&\, \frac{4}{5 \alpha \chi}(H-H_{\|}) -\frac{1}{5 \alpha}(2 H' + 3 H_{\|}'),\\
   \mathbb{X}^{(2)}_{2} 
                        \circeq&\, \frac{4}{3} (H_{\|}^2-H^2) - \frac{2}{3}(\dot H_{\|} - \dot H),\\
   \mathbb{X}^{(2)}_{3}
                        \circeq&\, \frac{2}{5 \alpha}(H' - H_{\|}') -\frac{4}{5 \alpha \chi}(H-H_{\|}),\\
    
    \mathbb{X}^{(3)}_{0} 
   \circeq &\, \frac{1}{3}(2\ddot H + \ddot H_{\|})+\frac{1}{15 \alpha^2}(2H''+3H_{\|}'') -\frac{\alpha'}{15\alpha^3}(2H'+3H_{\|}') +\frac{2}{3\alpha^2\chi}H_{\|}' \nonumber\\
        & +\frac{4}{15\alpha^2\chi^2}(H_{\|}-H)\left(1-\frac{\chi \alpha'}{\alpha}\right)-\frac{1}{15}\dot H_{\|}(2H+33H_{\|}) -\frac{2}{15}\dot H(34H+H_{\|})\nonumber\\
        & +\frac{2}{15}(28H^3 + 2H^2H_{\|}+2HH_{\|}^2+13H_{\|}^3),\\
   
    \mathbb{X}^{(3)}_{1}
   \circeq & \, \frac{2}{5 \alpha} (2 \dot H' + 3\dot H_{\|}' - 8HH'-12H_{\|}H_{\|}') + \frac{16}{5\alpha \chi}(H^2 - H_{\|}^2) + \frac{8}{5\alpha \chi}(\dot H_{\|}-\dot H),\\
    
    \mathbb{X}^{(3)}_{2} 
\circeq & \,\frac{2}{3}(\ddot H_{\|} - \ddot H) + \frac{4 \alpha'}{21 \alpha^3 \chi}(H-H_{\|}) + \frac{44}{21 \alpha^2 \chi^2}(H-H_{\|}) -\frac{2\alpha'}{21\alpha^3}(H'+6H_{\|}') \nonumber \\ &+\frac{2}{21 \alpha^2 \chi} (5 H_{\|}' -12H') 
   + \frac{2}{21 \alpha^2}(H''+6H_{\|}'') + \frac{8}{21}(10 H_{\|}^3-11 H^3) + \frac{4 H H_{\|}}{21}(H+H_{\|})\nonumber \\
   & +\frac{2}{21}(50 H\dot H - \dot H H_{\|} - H \dot H_{\|} - 48 H_{\|}\dot H_{\|}),\\
   
    \mathbb{X}^{(3)}_{3} 
  \circeq & \, \frac{-4}{5\alpha}(\dot H' -\dot H_{\|}'+4H_{\|}H_{\|}' -4HH') +\frac{8}{5\alpha \chi}(\dot H -\dot H_{\|} -2(H^2-H_{\|}^2)),\\
 
    \mathbb{X}^{(3)}_{4}
  \circeq & \, \frac{8}{35}\left(2(H^2 - H_{\|}^2) -(\dot H - \dot H_{\|})-\frac{2}{\alpha^2 \chi^2}(4+\frac{\alpha'}{\alpha}\chi)\right)(H-H_{\|})    -\frac{8}{35\alpha^2}(H''-H_{\|}'') \nonumber\\& +\frac{8}{7 \alpha^2 \chi}(1+\frac{1}{5}\frac{\alpha'}{\alpha}\chi)(H'-H_{\|}'),\\

    \mathbb{Y}^{(1)}_{0} 
    \circeq & -\frac{2 \left(\alpha +\alpha ^3 \left(\chi ^2 (H-H_{\|})^2-1\right)+\chi  \left(\alpha ^3 \chi  \left(2
   \dot H +\dot H_{\|}\right)-2 \alpha '\right)\right)}{3 \alpha ^3
   \chi ^2},\\
   
    \mathbb{Y}^{(1)}_{1}
 \circeq & -\frac{4}{\alpha}H' - \frac{4}{\alpha \chi}(H- H_{\|}),\\
 
    \mathbb{Y}^{(1)}_{2} \circeq &\, \frac{2}{3}\left(\frac{1-\alpha^2}{\alpha^2 \chi^2}+\frac{1}{\alpha^2 \chi}\frac{\alpha'}{\alpha}\right) +\frac{2}{3}\left(H(H_{\|}-H)+H_{\|}^2 -H^2 +\dot H_{\|}-\dot H\right).
\end{align}

Note that these expressions hold only for an off-center observer. For the central case, $\chi_o = 0$ and $J=0$. The monopoles for the case of a central observer are
\begin{align}
   \mathbb{X}^{(1)}_{0}\circeq & \, H, \\
   \mathbb{X}^{(2)}_{0}\circeq & \,2 H^2 - \dot H -2H',\\
   \mathbb{X}^{(3)}_{0}\circeq & \, 6H^3 -7H\dot H + \ddot H +4\dot H' +3H'' -16HH',\\
   \mathbb{Y}^{(1)}_{0}\circeq & \, -2(\dot H + H^2 \Omega_{k} ).
\end{align}

The expressions for the multipoles of $\mathbb{X}^{(4)}$ and $\mathbb{Y}^{(2)}$, associated with the snap parameter, are less transparent, and we quote them in Appendix \ref{appendixsnap}. This choice is further motivated by the fact that the CC parameters for the linear perturbation of the FLRW metric \cite{Kalbouneh:2024}, which we use as a benchmark for comparison, were calculated only up to $O(z^3)$, involving CC up to the jerk. Therefore, we neglect the snap in the following analysis.

\section{Accuracy of the covariant cosmographic reconstruction}

The effectiveness of the CC approach in reconstructing the the luminosity distance for an off-center  LTB  observer is investigated using the  specific inhomogeneous model for the mass distribution already investigated  by \cite{Kalbouneh:2024}. This model assumes  that the local geometry $(z<0.1)$ is influenced by a single spherically symmetric over- or under-dense structure located at a finite distance from the observer.
Despite its analytical simplicity, the model is sufficiently flexible to reproduce the observed multipole structure of the expansion rate fluctuation field, as inferred from real data \cite{kalbouneh_marinoni_bel_2023,Kalbouneh:2024,Kalbouneh:2024yjj,paper5}. It thus serves as an ideal testbed for evaluating the capability of the cosmographic method to recover the exact, fully relativistic luminosity distance associated with the Lemaître–Tolman–Bondi (LTB) solution. An additional advantage of this setup is that the same model has previously been studied within the framework of linear perturbation theory (LPT) applied to a homogeneous and isotropic FLRW background \cite{Kalbouneh:2024}, where the spherical inhomogeneity is treated as a small density perturbation embedded in an otherwise smooth universe. This provides a valuable opportunity for a direct comparison between the linearized approximation and the exact, non-perturbative LTB solution which we will detail in \S \ref{linltb}.

Following \cite{Kalbouneh:2024} we model a spherically symmetric matter over/underdense  region,   as
\begin{align}\label{phys density}
    \rho_{m0}(\chi) &=\rho_\infty\left(1+\delta(\chi)\right) ,
\end{align}
where $\rho_\infty \equiv \rho_{m0}(\chi\to \infty) = \left(6 \pi G t_0^2 \right)^{-1}$ and where $\delta$, the density contrast,  is given by 
\begin{align}
    \delta(\chi)
    &= \delta_c\left(1+\left(\frac{\chi}{R_s}\right) ^2\right) ^{-\frac{3}{2}},
    \label{modelba}
\end{align}
and depends on two parameters, the central density $\delta_c$ and the over/underdensity scale $R_s$. 
The asymptotic density for $\chi \rightarrow \infty$ is chosen to match the smooth background FLRW model, specifically the Einstein–de Sitter (EdS) Universe—a flat, matter-dominated model with simple analytical behavior. In this model, the FLRW scale factor evolves as $a(t) = \left( {t}/{t_0} \right)^{2/3}$
and the background Hubble rate is given by $H(t) = {2}/({3t})$. As such, we adopt for our analysis an LTB model containing only matter and curvature which asymptotically merges into the EdS model. 
\begin{figure}[htbp]
    \centering
    \includegraphics[scale=0.45]{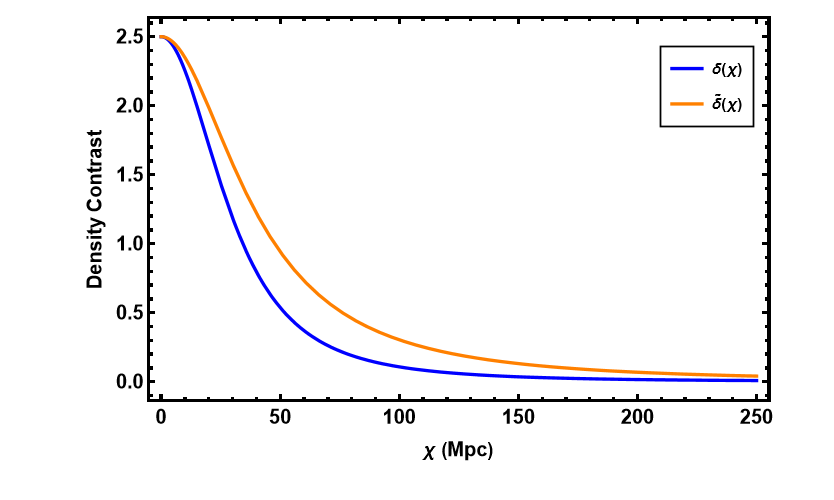}
    \includegraphics[scale=0.45]{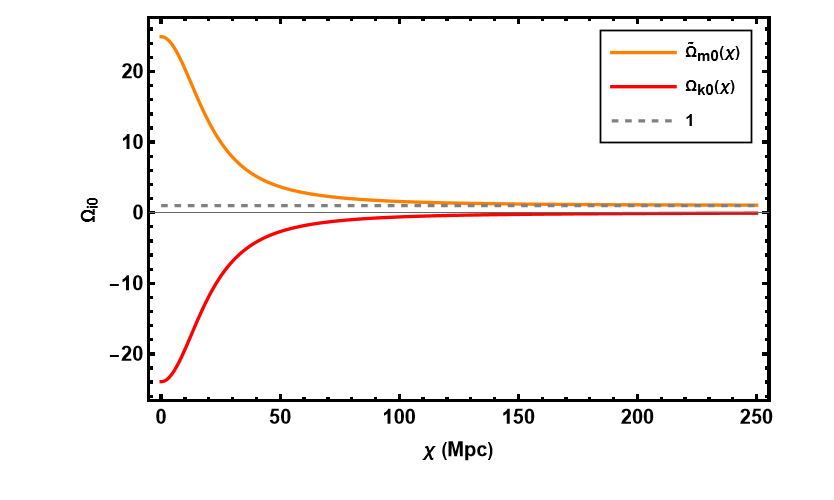}\\
    \includegraphics[scale=0.45]{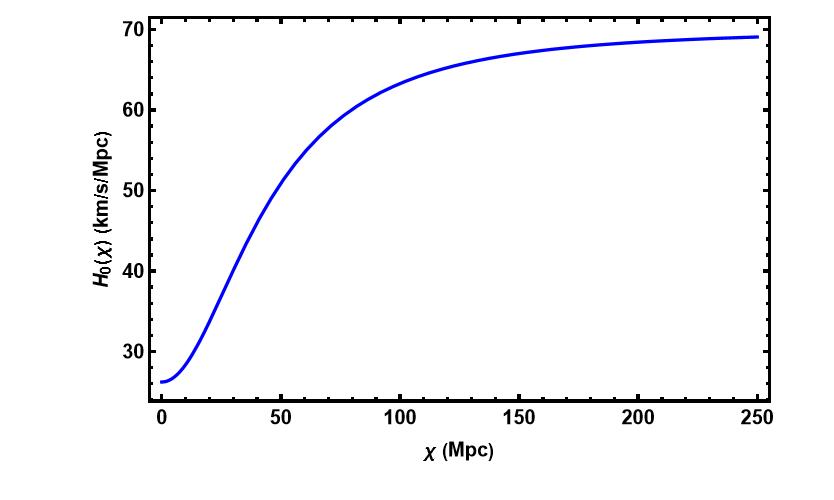}
  \caption{Radial scaling profiles of $\delta$, $\tilde \delta$, $\tilde{\Omega}_{m0}$, $\Omega_{k0}$ and $H_0$ are shown for $\delta_c = 2.5$ and $R_s=37.4$ Mpc.}
  \label{profiles}
\end{figure}
Therefore, the flat average matter density at present time can be obtained from (\ref{md}) as
\begin{align}\label{rhotil}
    \tilde\rho_{m0}(\chi) &= \frac{3}{\chi^3}\int_0^\chi \,\dd\bar{\chi} \   \rho_\infty \left(1+\delta\left(\bar{\chi}\right)\right) \bar{\chi}^2 \\
    &\equiv \rho_\infty \left( 1+ \tilde \delta(\chi)\right),
\end{align}
where we have defined $\tilde \delta$ as the density contrast associated with $\tilde \rho_{m0}$
\begin{align}
    \tilde \delta (\chi) &= 3\delta_c \left(\frac{R_s}{\chi}\right)^3 \left[ \sinh^{-1}\left(\frac{\chi}{R_s}\right) -\frac{1}{\sqrt{1+\left(\frac{R_s}{\chi}\right)^2}} \right].
\end{align}

By setting the current age of the universe  to be constant, independent of the radial coordinate, we obtain the following  constraint on the  Hubble profile at the present time by integrating (\ref{thexp}) over the scale factor $A$ from $0$ to $\chi$, which is given by 
\begin{align}\label{H0LTB}
H_0(\chi) &= 
\frac{1}{t_0 \left(\tilde \Omega_{m0}(\chi) - 1\right)} \left(\frac{\tilde \Omega_{m0}(\chi)}{\sqrt{\tilde \Omega_{m0}(\chi) - 1}} \tan^{-1}\left(\sqrt{\tilde \Omega_{m0}(\chi) - 1}\right) - 1\right).
\end{align}
Numerically $t_0$ is taken as the age of the asymptotic EdS, ${2}/({3 H_\infty})$, where $H_\infty = 70$ km/s/Mpc is the expansion rate in the EdS model. The radial scaling of the  characteristic LTB functions $\tilde\delta(\chi), H_0(\chi)$, $ \tilde{\Omega}_{m0}(\chi)$ and $\Omega_{k0}(\chi)$ are shown in Figure \ref{profiles}. They are obtained by modelling the LTB inhomogeneity by means of  the density profile  (\ref{modelba})  and the specific parameters of 
model $M1$ in \cite{Kalbouneh:2024} (from now onwards referred to as LTB$_{M1}$), specifically  $\delta_c=2.5$ and $R_s=37.4$ Mpc.
The radial Hubble profile at present time shows, 
as expected, that a large overdense region leads to a suppression of the local expansion rate in its vicinity, relative to the asymptotic value corresponding to a smooth FLRW background.

The impact on the estimation of $H_0$ is huge already for a moderately non-linear overdensity. As $H_0$ is a proxy for the monopole of $\Hbb_o$, which is the observable, a central underdensity can account for an increase in the locally measured monopole of the covariant cosmographic Hubble parameter  from an asymptotic value of $\mathbb{H}_0 = 68$ km/s/Mpc to a local value of $\mathbb{H}_0 = 73$ km/s/Mpc.  For a central observer ($\chi_o = 0$), this requires a central density contrast of $\delta_c \approx -0.22$. However, if the observer is located away from the center, a deeper underdensity is needed to produce the same effect. In the extreme case where the observer is $1.32 R_s$ from the center, the required central density contrast reaches the physical limit of $\delta_c = -1$, beyond which the density would become negative and thus unphysical. There is a degeneracy as such between the amplitude of density contrast and the position of the observer. This degeneracy can be broken by the quadrupole of $\Hbb_o$. For the case of an off-center observer with $\delta_c=-1$, the quadrupole of the covariant Hubble parameter is found to be $-7.37$ km/s/Mpc. However, it is found that for the case of an underdensity with an observer inside the void, linear perturbation theory (discussed in \S \ref{linltb}) provides a better estimation of the true LTB distance compared to cosmography.

\begin{figure}[htpb]
  \centering
  \includegraphics[width=0.55\textwidth]{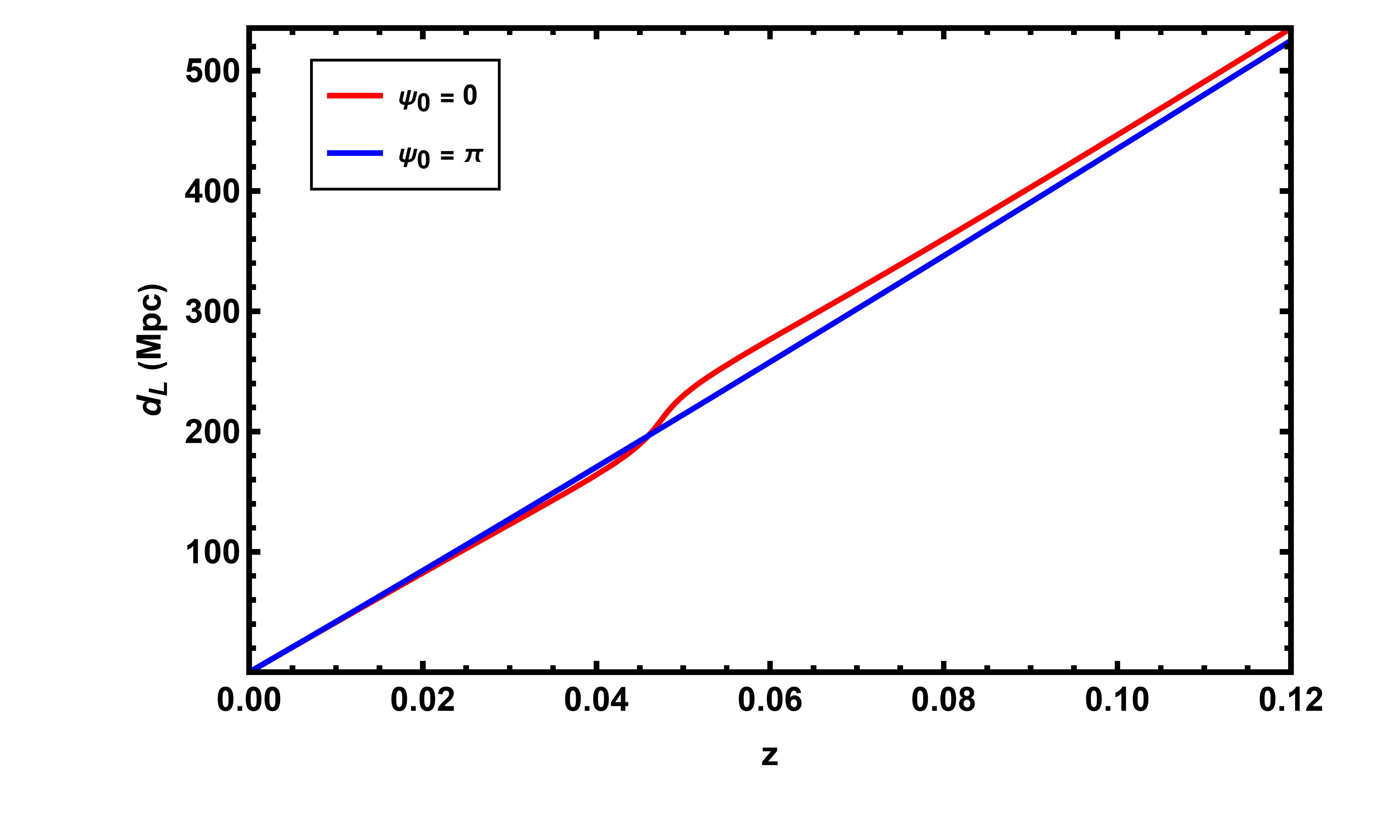}
  \caption{Luminosity distance for the  off-center observer in the  LTB$_{M1}$ model, calculated using eq. (\ref{dida}). 
  The scaling with redshift along two different line-of-sight directions is shown: towards the center of the density peak ($\psi=0$, solid red line) and in the antipodal direction ($\psi=\pi$, solid blue line).}
  \label{truedl}
\end{figure}

In Figure~\ref{truedl}, we show the luminosity distance \( d_L \) as a function of redshift \( z \), as measured by an off-center observer (200 Mpc from the center) in the LTB$_{M1}$ universe. This result is obtained from the exact relativistic solution computed via (\ref{dida}). The luminosity distance depends on the direction of the line of sight; we present its redshift scaling along two specific directions: toward the center of the mass overdensity ($  \psi = 0$) and in the opposite direction ($  \psi = \pi $).

The distance–redshift relation exhibits a characteristic elongated S-shaped trend along the observer’s line of sight, with a notable inflection near the location of the density peak. This feature mainly arises from directional variations in the observed redshift: sources located in front of the density peak have higher redshift than expected, while those beyond it have lower redshift, relative to a homogeneous background  with the same asymptotic density of the LTB model.

\begin{figure}[htbp]

  \subfloat[Towards the center ($\psi=0$).\label{fig:subfigA2}]{
    \includegraphics[width=0.51\textwidth]{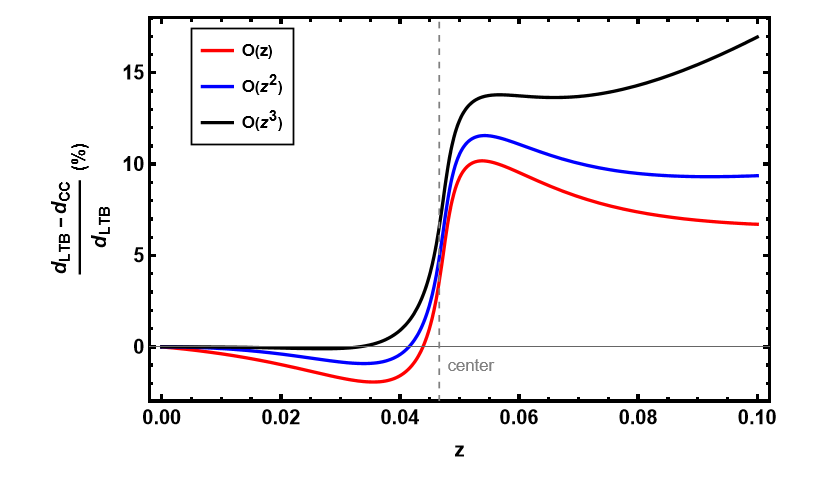}
  }
  \hfill
  \subfloat[Away from the center ($\psi=\pi$).\label{fig:subfigB3}]{
    \includegraphics[width=0.51\textwidth]{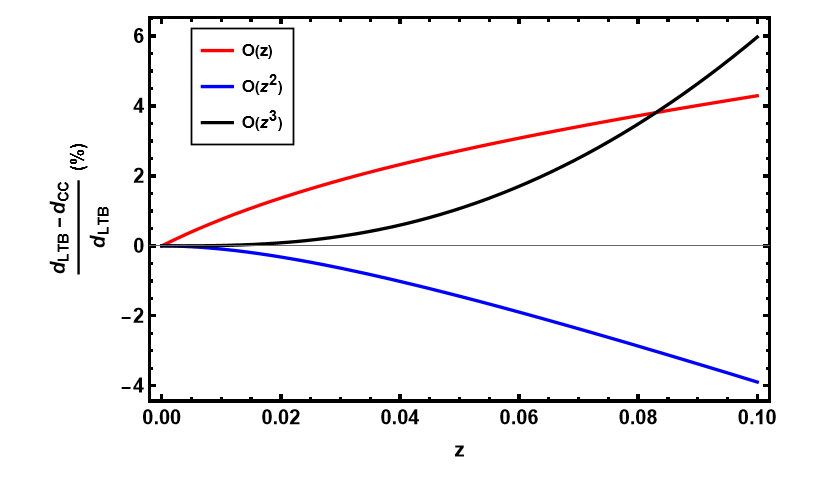}
  }
  \caption{The comparison between the exact luminosity distances for the LTB$_{M1}$  model and the distance reconstructed using the covariant cosmographic (CC) expansion in the two panels. The relative error in the cosmographic approximation is displayed for various expansion orders, along two directions: toward the center of the density peak ({\it left  panel}), and in the antipodal direction ({\it right  panel}). The dashed vertical line indicates the center of the mass overdensity.}
  \label{fig:order}
\end{figure}

The increased nonlinearity in the density field progressively degrades the accuracy of reconstructions based solely on low-order expansion terms. This is illustrated in 
Figure \ref{fig:order} where we show  the luminosity distance reconstructed    
using the cosmographic approach against the `exact'  unperturbed solution 
for the  off-center observer in the LTB$_{M1}$ metric. The relative discrepancy is shown along two line-of-sight directions ($\psi=0$ and $\psi=\pi$) and for different expansion orders in redshift.

\begin{figure}[htpb]
    \centering
    \includegraphics[scale=0.45]{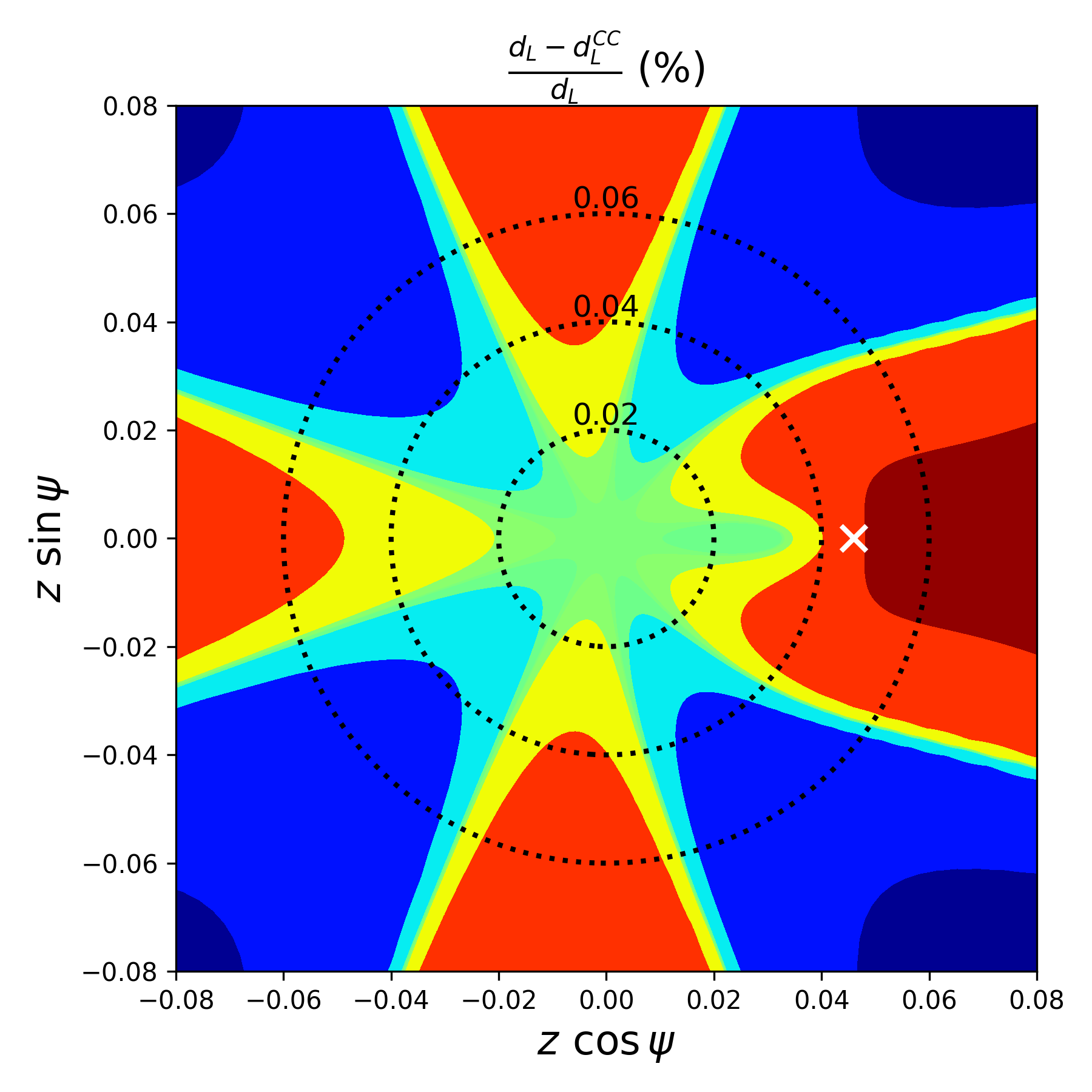}
    \includegraphics[scale=0.25]{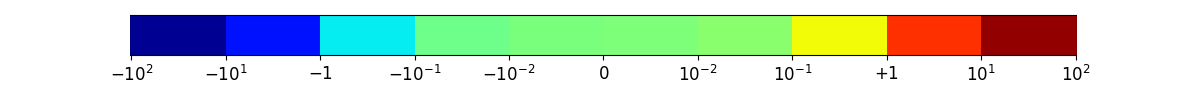}\\
    
    \caption{Relative error (in \%) in estimation of luminosity distance for an off-center observer using covariant cosmography up to $\mathcal{O}(z^3)$. The white "X" marks the center of the overdensity which is at a distance of 200 Mpc away from the observer.}
    \label{fig:dirdep}
\end{figure}

The cosmographic expansion of \( d_L \), as given in (\ref{dlcc}), is highly accurate in the immediate vicinity of the observer and along any line of sight.  
Its precision improves systematically with the inclusion of higher-order terms. However, even at relatively low redshifts—around \( z \approx 0.05 \) in the case of the M1 model—its accuracy degrades significantly in the presence of a nearby large density peak (with inaccuracies larger than 10\%). In Figure \ref{fig:dirdep}, we showed the directional dependence on the error.

The dominant multipoles (see section 7 in \cite{Kalbouneh:2024yjj}) of the CC parameters for the LTB$_{M1}$ with $\delta_c=2.5$ and $R_s=37.4$ Mpc with the observer at $\chi_o=200$ Mpc from the center of the over-density are 
\begin{align}
    \mathbb{H}_0 &= 69.622 \,\,\, \text{km/s/Mpc} ,\hspace{5mm} \mathbb{H}_2= 2.4  \,\,\, \text{km/s/Mpc} ,\\
    \mathbb{Q}_0 &=0.515 ,\hspace{5mm} \mathbb{Q}_1=-0.624 ,\hspace{5mm} \mathbb{Q}_3=1.822,\\
    \mathbb{J}_0 &= -5.81,\hspace{5mm} \mathbb{J}_2= -62.17,\hspace{5mm} \mathbb{J}_4=133.2\,.
\end{align}

In fact, a truncated polynomial expansion at a low order is ineffective in capturing these S-shaped features caused by large inhomogeneities. It is therefore essential to know at what scales the covariant cosmographic expansion fails if it is to be successfully applied in the highly irregular regions of the local Universe. 
These findings confirm those of \cite{Kalbouneh:2024}, both qualitatively and quantitatively. As in \cite{Kalbouneh:2024}, in the direction of the overdensity, CC systematically overestimates the distances of nearby objects and underestimates those of more distant ones. Moreover, the amplitude of this imprecision is comparable to that inferred in \cite{Kalbouneh:2024}, where the true distance was estimated using linear perturbation theory. Nevertheless, the error is slightly but systematically smaller across all the redshifts investigated, compared to those obtained by comparing CC to LPT. This motivates the further analysis presented in §\ref{linltb}.

\begin{figure}[htpb]
    \centering
        \includegraphics[scale=0.45]{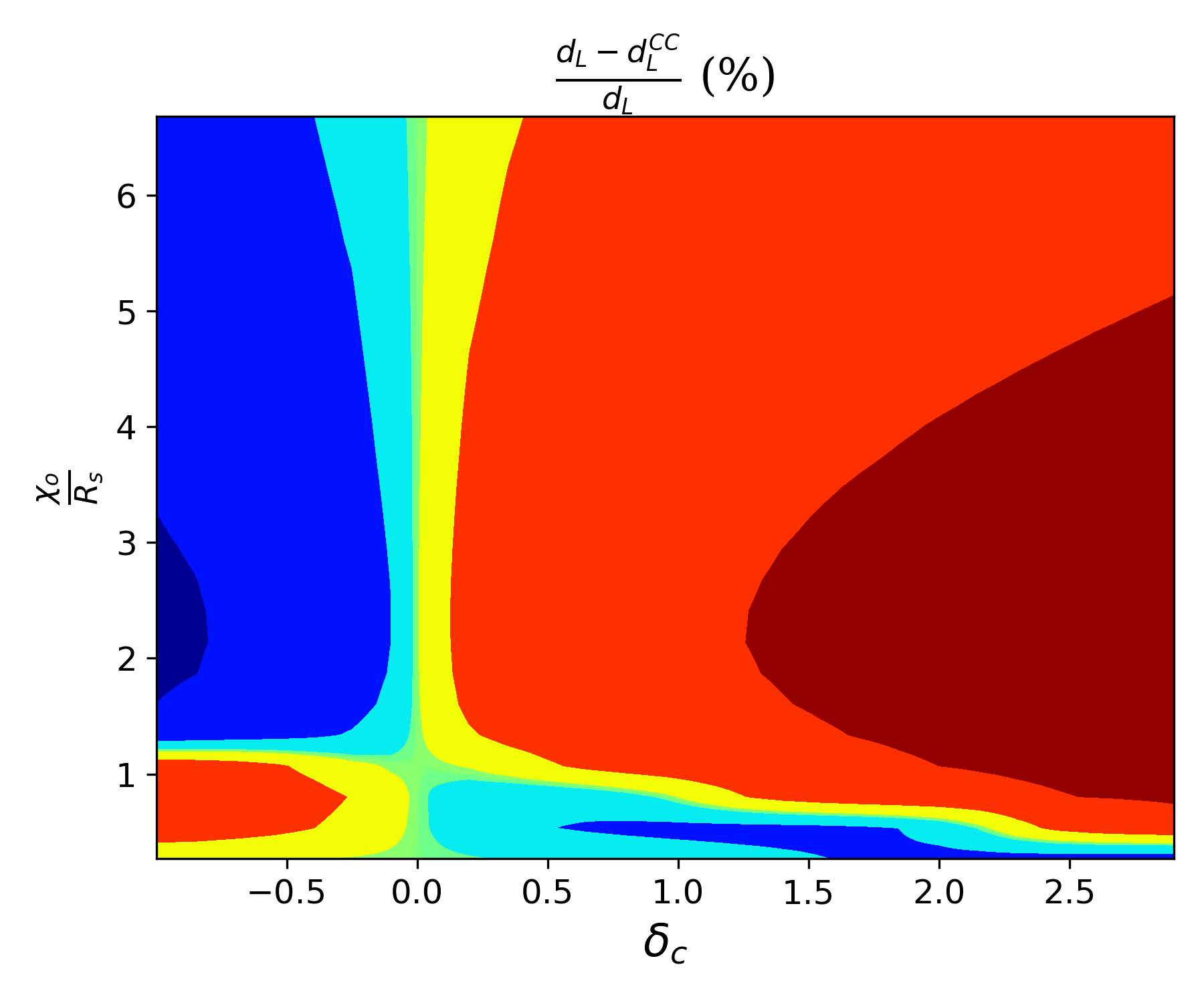}
        \includegraphics[scale=0.45]{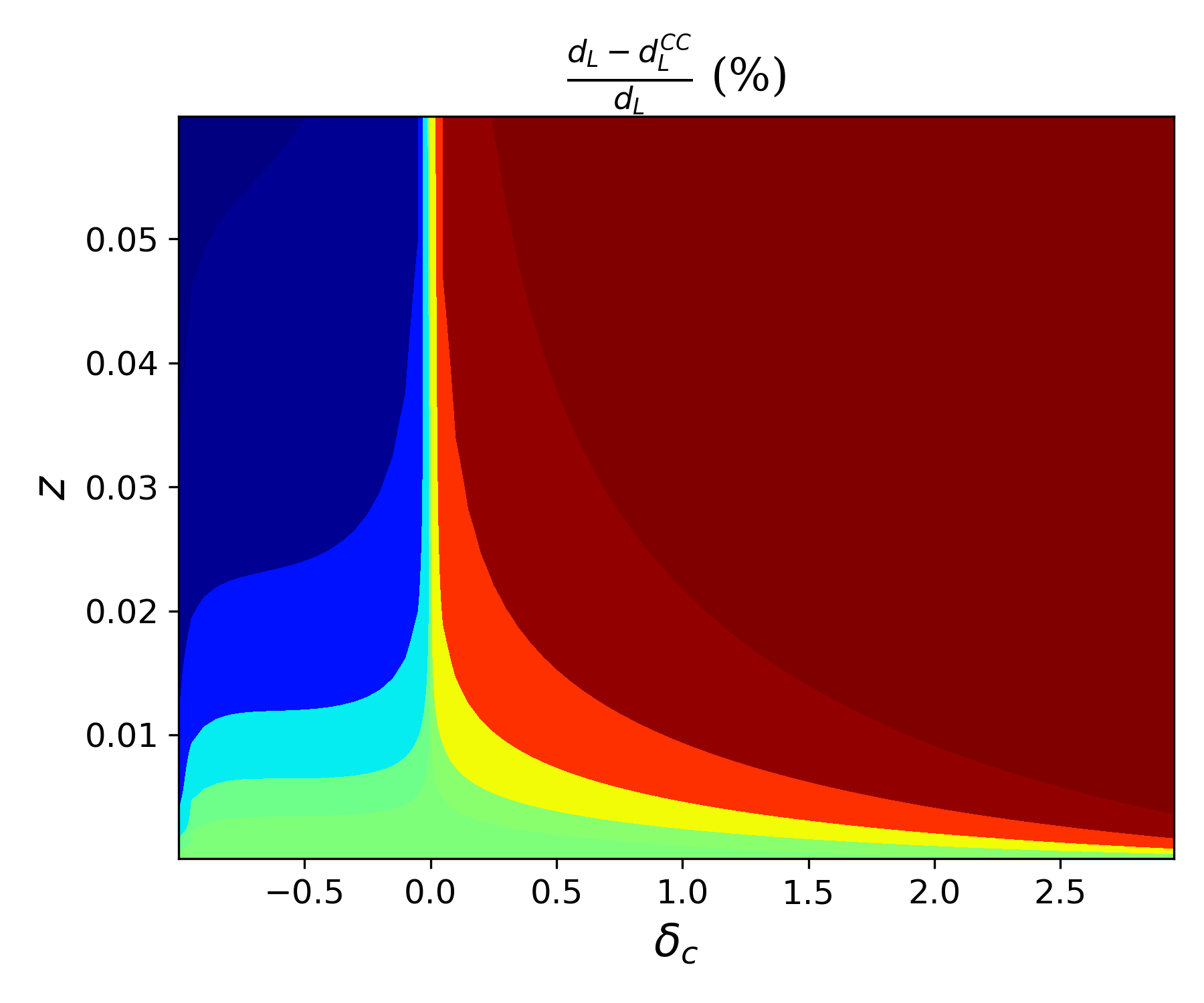}\\
    \includegraphics[scale=0.4]{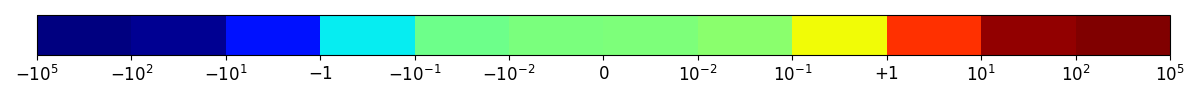}

    \caption{\small \textit{Left:} Relative error (in \%) in estimation of luminosity distance to the center of inhomogeneity for an off-center observer. Each point in the 2D parameter space corresponds to a specific value of the central density contrast $\delta_c$  and the observer's distance from the center, expressed in units of the characteristic size of the inhomogeneity $ R_s$, keeping $R_s$ fixed ($=37.4$ Mpc). \textit{Right:} Relative error (in \%) in the cosmographic (CC) reconstruction of the luminosity distance as a function of redshift and central density contrast for an observer located at the center of the inhomogeneity.}
        \label{fig:diag1}
\end{figure}

Figure~\ref{fig:diag1} (left) presents a diagnostic diagram in which, for each configuration of the inhomogeneous overdensity field—characterized by its central density amplitude $\delta_c$, size \( R_s \), and the observer's distance from the center $\chi_o$—we calculate the error incurred when approximating the exact relativistic solution for the luminosity distance to the center of the inhomogeneity with the third-order covariant cosmographic expansion. We consider the size of the structure $R_s$ to be $37.4$ Mpc. As the largest discrepancies are expected along the line of sight toward the overdensity, this is the reference direction used to evaluate the error.
As expected, if the gravitational source—regardless of its amplitude—is located far from the observer, the redshift range over which the cosmographic expansion remains accurate increases accordingly. This is due to the gentle slope of the density field at large distances from the peak. For a similar reason, even when the observer is located close to the peak, but the overdensity is mild, with a central peak \( |\delta_c| < 1 \), the covariant cosmographic approximation achieves better than 5\% accuracy locally (for $z<0.02$).

In Figure \ref{fig:diag1} (right), we find the rapid divergence of the cosmographic distance as a function of redshift with respect to the exact LTB luminosity distance for a central observer. This is expected due to the change of the shape of the structure which changes the derivatives. The higher the value of $\delta_c$, the faster the derivatives of the dynamical quantities change thereby losing the predictivity of the cosmographic expansion. Moreover, in this model, we find that the third order correction is more sensitive to positive $\delta_c$ than negative $\delta_c$. 

Covariant cosmographic parameters are formally defined as local derivatives of the distance – redshift relation, but in practice they are estimated by fitting over a finite redshift range, which stabilizes the inference and ensures physical relevance. This operational viewpoint also clarifies the minimal scale beyond which local data must be excluded to capture the large-scale geometry. The accuracy levels shown in Figure~\ref{fig:diag1} should be interpreted as the upper bound of the theoretical imprecision. The actual imprecision,  which is lower given the average non-local nature of the fitting procedure,  must be estimated by comparing the measured distances with those predicted by CC, using a maximum likelihood analysis.

\section{Comparision of CC in LTB with CC in Linear Perturbation Theory} \label{linltb}

It is of significant interest to compare not only how the covariant cosmographic formalism approximates the true relativistic distance to sources in an inhomogeneous LTB metric, but also how this underlying metric itself could be reconstructed using approximate perturbative approaches—particularly linear perturbation theory (LPT)—within the FLRW framework. 

Such a comparison enables us to shed light on the potential biases in cosmographic parameters that may arise when observational data are theoretically interpreted within a perturbed FLRW framework, as done for example by \cite{Kalbouneh:2024, Koksbang_2025},
rather than the true inhomogeneous relativistic geometry.

The starting point for this investigation is to ask whether the multipoles of the covariant cosmographic parameters obtained by an off-center observer in the LTB metric coincide with the multipoles obtained for a similar inhomogeneous matter density profile assumed to generate metric perturbations (in the Newtonian gauge) of the type  

\begin{equation}
\mathrm{d}s^2=-\left(1+2\Phi\right)\mathrm{d}t^2+\bar a^2(t)(1-2\Phi)\left[\mathrm{d}r^2+ r^2(\mathrm{d}\theta^2+\sin^2\theta \,\mathrm{d} \phi^2)\right],
\label{metric1}
\end{equation}
where $\Phi(r)$ is the time-independent gravitational potential generated by the  spherically symmetric structure. 

This, as we will see, leads to the more general question of assessing in which regimes the linearly perturbed line element (\ref{metric1}) provides a satisfactory description of the LTB line element (\ref{dsltb}).

LPT stems from the assumption that the amplitude of the matter overdensity perturbations is  small, i.e. in our case,  $|\delta_c| \ll 1.$ 
To compare LTB and  LPT predictions, we start by identifying the suitable LTB quantity that can be split into a background plus perturbation values and thus act as the relevant variable which  controls the region of convergence/ divergence between LPT and LTB predictions. Indeed if $|\delta_c| \ll 1$ then (\ref{oh}) can be expanded at leading order as 
\begin{eqnarray}\label{omegatil}
    \tilde\Omega_{m0}(\chi) &\approx 1 + \epsilon(\chi), 
\end{eqnarray}
where,  in our specific case, $\tilde{\Omega}_{m0}(\chi\rightarrow\infty)=1.$
By (\ref{H0LTB}), this also implies that the transverse expansion rate profile at present time   is
\begin{align}\label{H0approx}
    H_0(\chi) &\approx H_\infty \left(1-\frac{1}{5}\epsilon(\chi) \right),
\end{align}
where $ H_\infty \equiv 2/(3 t_0).$ The fact that $\epsilon $ is a good expansion parameter is verified by noticing that 
using (\ref{omegatil}), (\ref{rhotil}) and (\ref{H0approx}) in (\ref{oh}), we get
\begin{align}\label{phi-delta}
    \epsilon(\chi) &= \frac{5}{3} \tilde \delta(\chi).
\end{align}

We calculate the analytical expressions for the nonzero multipoles of the covariant cosmographic parameters (see (\ref{cc1})-(\ref{cc4})), assuming furthermore that the following ratio is small: 
\begin{align}
 \frac{H'_0 (\chi_o) \chi_o}{H_0(\chi_o)} 
    = \frac{\left(\frac{-2}{15 t_0}\right) \epsilon'(\chi_o)\chi_o}{\left(\frac{2}{3 t_0}\right)\left( 1+\frac{\epsilon(\chi_o)}{5}\right) }
     \approx -\frac{1}{5} \epsilon'(\chi_o) \chi_o 
    = - \frac{1}{3}\chi_o \tilde\delta'(\chi_o)\ll1,
\end{align}
where in the last equality, we have used (\ref{phi-delta}).

To facilitate comparison between the LTB multipoles and the LPT multipoles presented in \cite{Kalbouneh:2024} (Appendix B), we standardize the notation by introducing the dimensionless parameters  $\xi_0 \equiv 
R_s/\chi_o$ and $\xi_H \equiv R_s/R_H= H_\infty R_s$. Neglecting $\mathcal{O}(\delta_c^2)$ and higher-order terms, we are left with\footnote{Note that the expression for the quadrupole of the Hubble parameter ($\mathbb{H}_2$) in \cite{Kalbouneh:2024} is missing the square of $\xi_o$ in the denominator.}
\begingroup
\allowdisplaybreaks
\begin{align}
\mathbb{H}_0 &=
H_\infty\left[1-\frac{1}{3}\delta_c \: \xi_o\left(\frac{\xi_o^2}{(1+\xi_o^2)^{3/2}}\right)\right], \label{monoho}
\\
\mathbb{H}_2 &=
\frac{2\delta_c H_\infty \xi_o^3}{3(1+\xi_o^2)^{3/2}}\left[-4-3\xi_o^2+3(1+\xi_o^2)^{3/2}\csch^{-1}(\xi_o)\right],
\\
\mathbb{Q}_0 &=q_{\infty}+\frac{5\delta_c \xi_o^3}{6 \left(1+\xi_o^2\right)^{3/2}},\label{appb3}
\\
\mathbb{Q}_1 &=-\frac{9\delta_c \xi_o^4}{5 \xi_H (1+\xi_o^2)^{5/2}},
\\
\mathbb{Q}_2 &=-\frac{5}{2}\frac{\mathbb{H}_2}{H_\infty} \,,
\\
\mathbb{Q}_3&=\frac{2\delta_c \xi_o^4}{5 \xi_H (1+\xi_o^2)^{5/2}}\left[15(1+\xi_o^2)^{5/2}\csch^{-1}(\xi_o)-23-35\xi_o^2-15\xi_o^4\right],
\\
\mathbb{R}_0 &=1-2\mathbb{Q}_0\,,
\\
\mathbb{R}_1 &=\mathbb{Q}_1 \,,
\\
\mathbb{R}_2 &=-\frac{\mathbb{Q}_2}{5},
\\
\mathbb{R}_3 &=\mathbb{Q}_3 \,,
\\
\mathbb{J}_0 &=j_\infty-\frac{\delta_c \xi_o^3}{15 \xi_H^{2} (1+\xi_o^2)^{7/2}}\left[9\xi_o^2(2-3\xi_o^2)-25 \xi_H^2 (1+\xi_o^2)^2\right],
\\
\mathbb{J}_1&=-\mathbb{Q}_1\,,
\\
\mathbb{J}_2 &=- \frac{10\delta_c \xi_o^3}{21 \xi_H^{2} (1+\xi_o^2)^{7/2}}\left[18\xi_o^2-7(1+\xi_o^2)^2 (4+3\xi_o^2) \xi_H^2+21 \xi_H^2 (1+\xi_o^2)^{7/2}\csch^{-1}(\xi_o)\right],
\\
\mathbb{J}_3 &=-\mathbb{Q}_3\,,
\\
\mathbb{J}_4 &=\frac{8\delta_c \xi_o^5}{35 \xi_H^{2} (1+\xi_o^2)^{7/2}}\left[105(1+\xi_o^2)^{7/2}\csch^{-1}(\xi_o)-7\xi_o^2(58+50\xi_o^2+15\xi_o^4)-176\right],
\end{align}
\endgroup
where $q_\infty=0.5$ and $j_\infty=1$ are the deceleration and the jerk parameters respectively for the EdS background. All other multipoles vanish. Thus, we find that all multipoles—except for the Hubble monopole $\mathbb{H}_0$—exactly match the LPT multipoles obtained in \cite{Kalbouneh:2024}. The Hubble monopole 
differs from the value 
predicted in \cite{Kalbouneh:2024} by a term $(3/2)H_{\infty} \delta_c \xi_o\xi_{H}^2 \csch^{-1}(\xi_o)$.
This mismatch therefore calls for a more in-depth investigation (see \S \ref{monres}). Rather than merely linearizing the multipoles of the covariant cosmographic (CC) parameters, we pursue the broader goal of directly relating LTB and LPT observables. This approach not only helps identify the origin of the discrepancy in the Hubble monopole but also enables a more comprehensive comparison between LTB and LPT predictions—one that goes beyond the CC parameters and addresses the very notions of distance as inferred in these two frameworks.

\subsection{Linearisation of the LTB metric}\label{ltbline}

We begin by linearizing the LTB spacetime around an asymptotic FLRW background. In its native coordinate system, the linear LTB metric effectively operates in a gauge closely resembling the synchronous gauge, wherein the time coordinate corresponds to the proper time of comoving observers and the metric perturbation contains only spatial components. To establish consistency with LPT, we explicitly transform the linearized LTB perturbations into the synchronous gauge framework (for more details, see \cite{Ma_1995,spdust,Van_Acoleyen_2008}). Subsequently, to connect with cosmological observables (the covariant cosmographic parameters) and enable direct comparison with standard LPT results, we perform a gauge transformation from the synchronous to the Newtonian gauge. This transformation introduces scalar potentials $\Phi$ and $\Psi$ in the metric, which directly govern gravitational redshift and lensing observables. Crucially, while higher-order multipoles of the cosmographic expansion match across both gauges, the monopole term—dominated by the local Hubble flow—exhibits a gauge dependency, thereby revealing the necessity of proper gauge treatment when interpreting cosmological parameters in inhomogeneous cosmologies. 

Let us assume that the transverse scale factor $A$ of the LTB metric in (\ref{dsltb}) can be written in terms of the scale factor $\bar a$ of the asymptotic FLRW as 
\begin{align}
    A(t,\chi) &\approx \bar a(t) \chi\left(1+\frac{\delta a (t,\chi)}{\bar a(t)}\right) ,\,\,\,\quad \Bigg|\frac{\delta a (t,\chi)}{\bar a(t)}\Bigg|\ll1,
\end{align}
with $\bar a(t_0)=1$ and $\delta a (t_0,\chi)=0$ in order to maintain the normalization $A(t_0,\chi)=\chi$. Also, $|k|\ll1$ gives
\begin{align}{\label{small curv}}
    \alpha(t,\chi) &= \frac{A'}{\sqrt{1-k}} \approx \bar a(t) \left(1+ \frac{\delta a (t,\chi)}{\bar a(t)}+ \frac{\delta a'(t,\chi)}{\bar a(t)}\chi+\frac{k(\chi)}{2}\right).
\end{align}
Rewriting the LTB metric (\ref{dsltb}) using these linearised parameters and defining a conformal time $\eta_y$, such that, $\dd t=\bar a \,
\dd\eta_y$ , we get,
\begin{align}
    \dd s^2 &= \bar a^2 \left[-\dd\eta_y^2 + \left(1+2\left(\frac{k}{2} + \frac{\delta a}{\bar a} + \chi \frac{\delta a'}{\bar a}\right)\right)\dd\chi^2 + \chi^2\left(1+2\frac{\delta a}{\bar a}\right)\dd\Omega^2\right].
    \label{ltbml}
\end{align}
In the Cartesian coordinates $y^i$, this can be written as
\begin{align}\label{ltbcart}
    \dd s^2 &= \bar a^2 \left[-\dd\eta_y^2 +\left(1+2\frac{\delta a}{\bar a} \right)\delta_{ij} + \left(k+2\chi \frac{\delta a'}{\bar a} \right)\partial_i \chi \partial_j \chi \right]\dd y^i \dd y^j.
\end{align}
This  bears a resemblance to the scalar-perturbed FLRW metric in the synchronous gauge:
\begin{equation}
    \dd s^2 = \bar a^2 \left[-\dd\eta_y^2 +\left(\delta_{ij} + h_{ij}\right)  \dd y^i \dd y^j\right],\label{met2}
\end{equation}
where the perturbations $|h_{ij}|\ll1$ can be decomposed into trace and traceless components as  
\begin{align} \label{hij}
    h_{ij} &= \frac{h(t,\chi)}{3} \delta_{ij} + \left(\partial_i \partial_j - \frac{1}{3} \delta_{ij} \nabla^2\right) Q(t,\chi),\\
    &= \left(\frac{h}{3} -\frac{1}{3}\nabla^2Q + \frac{Q'}{\chi} \right)\delta_{ij} + \left(Q''-\frac{1}{\chi}Q'\right)\partial_i \chi \partial_j \chi ,
\end{align}
with $Q(\eta_y, \mathbf{y})$ and $h(\eta_y, \mathbf{y})$ being two scalar functions. Here, $\bar{a}(\eta_y)$ denotes the scale factor of the smooth, homogeneous background universe. Substituting this in (\ref{met2}), we get
\begin{align}\label{syncart}
    \dd s^2 &= \bar a^2 \left[-\dd\eta_y^2 +\left( \left( 1+ \frac{h}{3} -\frac{1}{3}\nabla^2Q + \frac{Q'}{\chi} \right)\delta_{ij} + \left(Q''-\frac{1}{\chi}Q' \right)\partial_i\chi \partial_j\chi \right)\dd y^i \dd y^j\right].
\end{align}
 
Both metrics (\ref{ltbcart}) and (\ref{syncart}) involve two perturbation parameters. As such, their respective characteristic quantities can be matched under the assumption that the time coordinate (cosmic time $t$), the conformal time ($\eta_y$) and the spatial coordinates $y^i$ (comoving with matter) are the same in both descriptions. Fixing the coordinates in this way fixes the synchronous gauge completely. This gives us the two  matching conditions,
\begin{align}
    2 \frac{\delta a}{\bar a} &= \frac{h}{3} -\frac{1}{3}\nabla^2Q + \frac{Q'}{\chi} ,\label{match1}\\
    k+2\chi \frac{\delta a'}{\bar a} &= Q''-\frac{1}{\chi}Q'. \label{match2}
\end{align}

\subsection{Dictionary between linearised LTB (LLTB) and LPT in conformal Newtonian gauge (CNG)}

We are now left with the task of implementing the gauge transformation from the synchronous gauge to the conformal Newtonian gauge, in order to understand how the characteristic parameters of the LLTB model—particularly the longitudinal and transverse scale factors, as well as the curvature parameter—relate to the scale factor and gravitational potential appearing in the Newtonian gauge formulation of the linearly perturbed FLRW metric.
This dictionary will in the end allow us also to consistently predict the correct 
amplitude for the monopole of the covariant Hubble parameter.

Consider  the perturbed flat FLRW  metric in the conformal Newtonian gauge $x^{\mu}=(\eta, \textbf{x})$,
\begin{equation}\label{cng}
    \dd s^2 = \bar a^2(\eta)\left[ -\left(1+2\Psi(\eta,\bold x)\right)\dd\eta^2 + \left(1-2\Phi(\eta,\bold x)\right)\dd x^i \dd x_i \right],
\end{equation}
as well as 
the infinitesimal  gauge transformation  $x^\mu \to  y^\mu = x^\mu + d^\mu (x^\nu)$,
where $d^{\mu}=(T, L^{,i})$ with $T(\eta,\bold x)$ and $L(\eta,\bold x)$ being two arbitrary scalar gauge fields. 

The gauge fields $T, L$ that transform the conformal Newtonian gauge metric (\ref{cng}) into the synchronous form (\ref{syncart}) are solutions of the set of equations
\begin{align}\label{gauge eq}
0&=\Psi -\pdv{T}{\eta} -\frac{1}{\bar a}\dv{\bar a}{\eta}T,\\
0&=T - \pdv{L}{\eta}, \label{gauge eq2} \\
-\frac{h}{6} &=\Phi +\frac{1}{3}\nabla^2L +\frac{1}{\bar a}\dv{\bar a}{\eta}T,  \label{gauge eq3} \\
Q&=-2L. \label{sp1}
\end{align}
Assuming no anisotropic stress, we have $\Psi=\Phi$. This sets a constraint on $h$ and $L$ given by
\begin{align}\label{h}
    h &= -6\left( \pdv[2]{L}{\eta} + 2\frac{1}{\bar a}\dv{\bar a}{\eta}\pdv{L}{\eta} + \frac{1}{3}\nabla^2 L \right). 
\end{align}
Substituting $T$ from (\ref{gauge eq2}) in (\ref{gauge eq}) and solving for $L$, we get
\begin{align}\label{generalGT}
    L(\eta,\bold x) &= \int_0^\eta \dd\eta'\, \left(\frac{1}{\bar a}\int_0^{\eta'} \dd{\eta''} \,  \bar a \Phi(\eta'',\bold x)\right) + C_2(\bold x) \int_0^\eta   \frac{\dd{\eta'}}{\bar a} + C_1(\bold x).
\end{align}
In the synchronous gauge, we want the observer to be comoving with matter. The transformation of the 4-velocity of the observer puts a constraint on the integration constant $C_2(\textbf{x})$ given by
\begin{align}
    u_N^\mu &= \pdv{x^\mu}{ y ^\nu}u_S^\nu ,
\end{align}
where 
\begin{align}
   u_N^\mu(\eta,\textbf{x})
    &= \begin{pmatrix}  
        \dfrac{1 - \Phi(\eta,\textbf{x})}{\bar a(\eta)} \\\\
        \dfrac{\textbf{v}(\eta,\textbf{x})}{\bar a(\eta)} 
    \end{pmatrix}
    \qquad\qquad 
    \text{and} 
    \qquad\qquad 
    u_S^\mu(\eta_y)
    = \begin{pmatrix}  
        \dfrac{1}{\bar a(\eta_y)} \\\\ 
        \textbf{0} 
    \end{pmatrix}.
\end{align}
Then we get the velocity constraint
\begin{align}
    v^i  =& \pdv{x^i}{ \eta_y} \nonumber\\ 
    \approx &  -\frac{\partial}{\partial \eta}(\partial^i L)  \nonumber\\
               = & - \frac{1}{\bar a}\int_0^\eta \dd{\eta'} \bar a \partial^i \Phi  - \frac{1}{\bar a}\partial^i C_2.
\end{align}

If we take the velocity field to be generated due to the gradient of the potential $\Phi$ alone, then from the spatial part of the conservation of the of the energy momentum tensor in linear perturbation theory, we get
\begin{align}
    v^i &=- \frac{1}{\bar a}\int_0^\eta \dd{\eta'} \bar a \partial^i \Phi.
\end{align}

We deduce that $   C_2 = \text{const.}$ We choose $C_2 = 0$, since the metric must converge to the unperturbed FLRW solution at spatial infinity, where the potential $\Phi$ 
vanishes. Note that this assumption holds only in the Newtonian regime of linear perturbation theory, where the perturbations are smaller than the Hubble scale, 
and we consider their evolution to be ruled only by the growing mode.
This assumption holds here as the scale is set by the curvature constraint (\ref{small curv}). Thus, the synchronous gauge perturbations can be written in terms of the CNG perturbation $\Phi(\eta,\textbf{x})$ using (\ref{gauge eq3}) and (\ref{sp1}) as 
\begin{align}
h(\eta, \textbf{x}) &= -6\left[ \Phi + \frac{1}{\bar a^2}\dv{\bar a}{\eta}\int_0^\eta \dd{\eta'}\,\bar a\Phi   +\frac{1}{3}\int_0^\eta \dd{\eta'}\left(\frac{1}{\bar a}\int_0^{\eta'} \dd{\eta''}\,\bar a\nabla^2\Phi\right) +\frac{1}{3}\nabla^2C_1 \right],\\
    Q(\eta,\bold x) &= -2C_1 -2\int_0^\eta \dd{\eta'}\left( \frac{1}{\bar a} \int_0^{\eta'} \dd{\eta''} \,\bar a \Phi\right).
\end{align}

We are now in a position to establish the connection between the relevant LTB quantities---such as the scale factors and curvature---and the LPT quantities in the CNG, namely the gravitational potential $\Phi$. Using (\ref{match1}) and (\ref{match2}),
\begin{align}
    \frac{\delta a(\eta,\bold x)}{\bar a(\eta)} &= -\Phi -\frac{1}{\bar a^2} \dv{\bar a}{\eta} \int_0^\eta \dd{\eta'} \,\bar a\Phi - \frac{C_1'}{\chi} - \frac{1}{\chi} \int_0^\eta \dd{\eta'}\left( \frac{1}{\bar a}\int_0^{\eta'} \dd{\eta''} \,\bar a\Phi'\right),\label{matchf}\\
    k(\bold x) &= 2\chi \Phi' + \frac{2\chi}{\bar a^2}\dv{\bar a}{\eta}\int_0^\eta \dd{\eta'}\,\bar a\Phi' . \label{matchk}
\end{align}
The curvature is independent of the residual gauge $C_1$. It should also be time independent. Therefore, the second term in the RHS of (\ref{matchk}) should be time independent. We can fix $C_1$ from the constraint that $\delta a(\eta_{y0},\bold y) \approx \delta a(\eta_{0},\bold x)= 0$.
\begin{align}
  C_1' &=    -\chi \left(\Phi +\left(\frac{1}{\bar a^2} \dv{\bar a}{\eta}\right)\Bigg|_{\eta_0} \int_0^{\eta_0} \dd{\eta'} \,\bar a\Phi+ \frac{1}{\chi} \int_0^{\eta_0} \dd{\eta'}\left( \frac{1}{\bar a}\int_0^{\eta'} \dd{\eta''} \,\bar a\Phi'\right)\right).
\end{align}
Using this in (\ref{matchf}), 
\begin{align}
    \frac{\delta a(\eta,\bold x) }{\bar a(\eta)} &= -\frac{1}{\chi} \int_{\eta_0}^{\eta}\dd{\eta'} \left( \frac{1}{\bar a}\int_0^{\eta'} \dd{\eta''} \,\bar a\Phi'\right) - \frac{1}{\bar a^2}\dv{\bar a}{\eta}\int_0^\eta \dd{\eta'}\,\bar a\Phi + \left(\frac{1}{\bar a^2} \dv{\bar a}{\eta}\right)\Bigg|_{\eta_0}\int_0^{\eta_0} \dd{\eta'}\,\bar a\Phi .  
\end{align}
For the case of an EdS background, $\bar a(\eta)\propto \eta^2$ and for a spherically symmetric potential $\Phi(\chi)$ which is independent of time,
\begin{align}
   \frac{\delta a(\eta_y,\chi) }{\bar a(\eta_y)} &= -\frac{2\Phi'(\chi)}{3H_{\infty}^2\chi}\left(\bar a(\eta_y)-1\right) = \frac{v(\chi)}{\chi H_\infty}\left(\bar a(\eta_y) -1\right),\label{feds}\\
     k(\chi)  &= \frac{10}{3}\Phi'(\chi) \chi = -5 H_\infty \chi v(\chi) \label{keds},
\end{align}
where $v(\chi)$ is the radial component of the velocity at $\eta_{y0}$. Thus, we get the radial and the transverse scale factors of the linear LTB,
\begin{align}
    A(t, \chi) &=\bar a(t) \chi \left( 1-\frac{2\Phi'(\chi)}{3 H_\infty^2 \chi}\left(\bar a(t)-1\right)\right),\label{Alin}\\
   \alpha(t, \chi) &=\bar a(t) \left( 1-\frac{2\Phi'(\chi)}{3 H_\infty^2}\left(\bar a(t)-1\right) +\frac{10}{3}\chi \Phi'(\chi)\right).\label{alphalin}
\end{align}

As expected, the curvature function vanishes at infinity, and the LTB scale factor becomes a separable function of spatial and temporal coordinates: \( A = \bar{a}(t)\chi\).
Note that, if the Newtonian-gauge perturbation $\Phi$ is expressed in terms of the integral of the curvature function $k$ (see \eqref{keds}), then $\Phi$ coincides with the growing mode given in (3.21) of \cite{Van_Acoleyen_2008}, where the linearized LTB approximation was explicitly assumed (density contrast $\delta < 1$ and peculiar velocity $v/c \ll 1$). One may also verify directly that the functions $A$ and $k$ satisfy the LTB Friedmann-like equations at linear order, reducing in this limit to (2.13) and (2.16) of \cite{Van_Acoleyen_2008}. Our approach, however, has the advantage of explicitly describing the coordinate transformation that relates the linearized LTB parameters to those of the perturbed FLRW framework (see (\ref{Alin}) and (\ref{alphalin})).

\begin{figure}[htpb]
    \centering
        \includegraphics[scale=0.45]{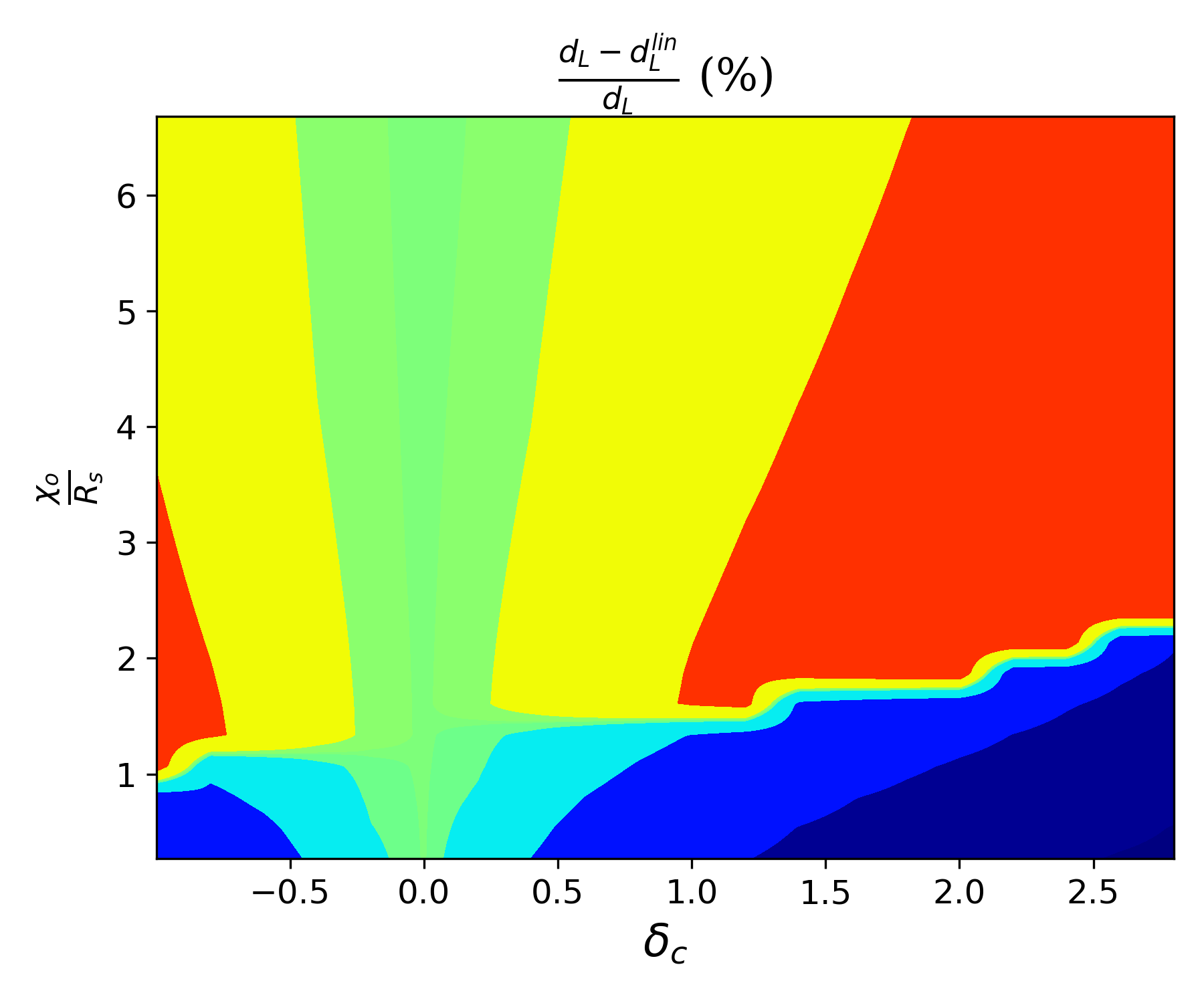}
        \includegraphics[scale=0.45]{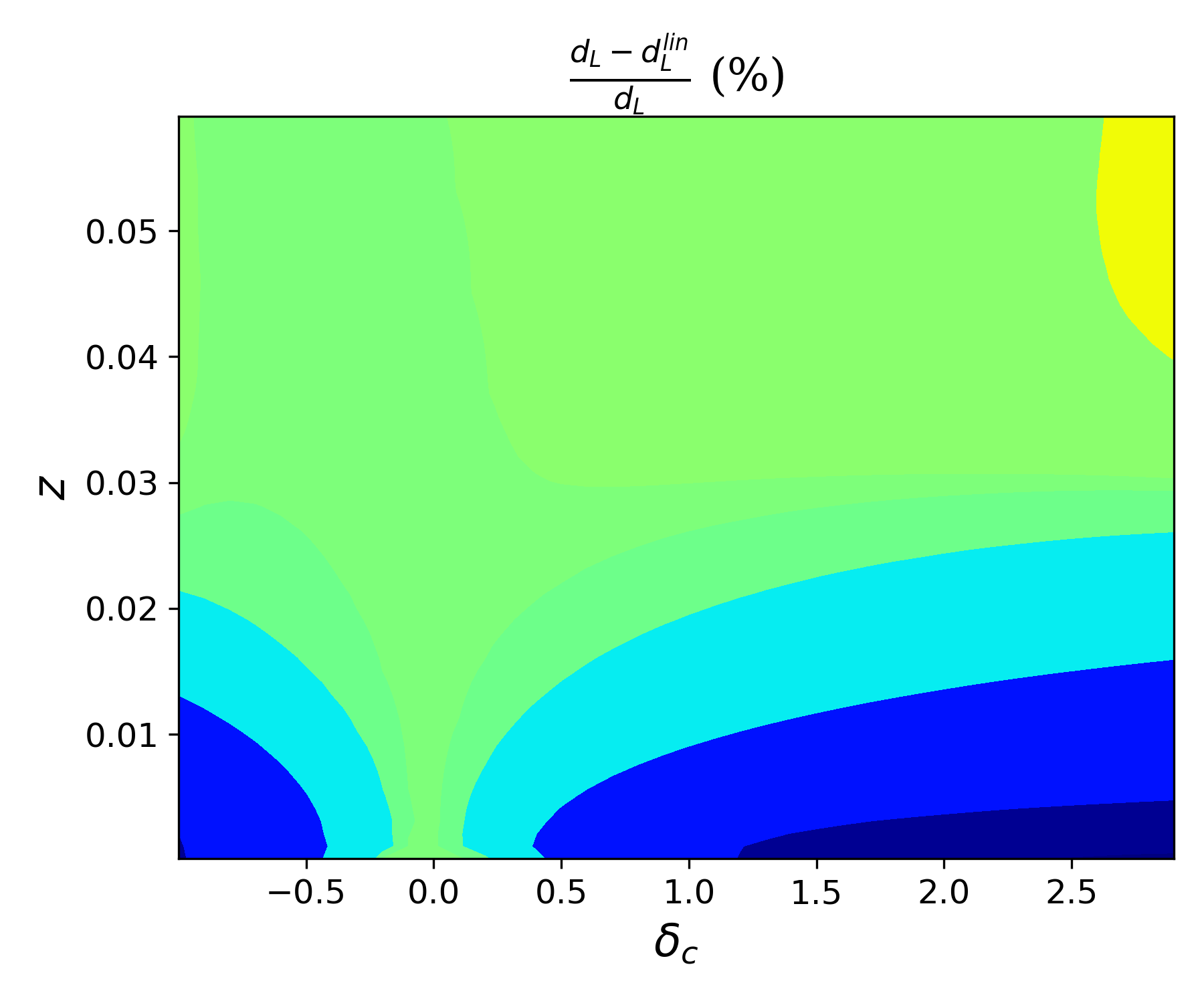}\\
    \includegraphics[scale=0.35]{fig-6c.png}

    \caption{\small\textit{Left:} Maximum relative error (in \%) in estimation of luminosity distance towards the center of inhomogeneity for an off-center observer. Each point in the 2D parameter space corresponds to a specific value of the central density contrast $\delta_c$  and the observer's distance from the center, expressed in units of the characteristic size of the inhomogeneity $ R_s$, keeping $R_s$ fixed ($=37.4$ Mpc). \textit{Right:} Relative error (in \%) in the luminosity distance between LTB and LLTB as a function of redshift and central density contrast for an observer located at the center of the inhomogeneity.}
    \label{fig:diag2}
\end{figure}

The luminosity distance calculated using both the full LTB metric and its linear approximation (see \ref{ltbml}) is shown in Figure~\ref{fig:diag2}. The latter coincides with expression (5.13) in \cite{Kalbouneh:2024}, which was derived using LPT within the standard cosmological model.

In Figure \ref{fig:diag2} (left), we find that overall the maximum relative error between LTB and LLTB luminosity distances decreases as we place the observer further from the center, i.e. in regions  where the density contrast is small. For the same reason, it decreases also when the observer is placed at the center of the inhomogeneity and the emitting source at progressively  higher redshift for a given $\delta_c$. Interestingly, linear perturbation theory proves more effective at reproducing the exact luminosity distance in the case of an underdensity, regardless of how empty the region is. In other words, it performs slightly better for underdensities than for overdensities.  Although LPT treates under- and overdensities symmetrically, gravitational collapse accelerates overdense regions into the nonlinear regime sooner. Underdensities, or voids, simply expand and get smoother; they don’t collapse, making the linear approximation more robust. 

Inspecting Figure~\ref{fig:diag2} (right), we see that model \( M1 \), used in \cite{Kalbouneh:2024} and characterized by a central density contrast $ \delta_c = 2.5$ , is still reasonably well described by linear perturbation theory for an observer at 200 Mpc away from the center. The maximum relative deviation of the LPT luminosity distance from the full relativistic LTB solution remains below 10\% in this case. In contrast, LPT becomes a poor approximation—exhibiting inaccuracies exceeding 10\%—even for modest central density peaks with \( \delta_c \sim 3 \), when the observer is located at less than twice the characteristic scale of the perturbation.

 If we increase $|\delta_c|$, the error in distance at the center increases non-linearly. Moreover, an issue that occurs at $\delta_c>3$ is that the linearized luminosity distance becomes multivalued with $z$ very close to the center, owing to the rapid transition of the velocity of the observer. This is not observed for the LTB case.

\subsection{Relating the multipoles of CC parameters in LLTB and LPT frameworks}\label{monres}

We now have the necessary ingredients to  properly relate the multipoles of the CC parameters as predicted by an off-center LTB observer to those derived from describing the same matter inhomogeneity within the framework of LPT. This ultimately boils down to identifying and resolving the origin of the $\mathbb{H}_0$ monopole mismatch.

In the following subsection, we will work with the LLTB asymptotic EdS limit and time independent potential $\Phi$ so as to recover the result of \cite{Kalbouneh:2024}. Having established the relation between the linear LTB parameters and those characterizing the LPT metric in the synchronous gauge, we can express the trace of the matter expansion tensor $\Theta_{\mu \nu}$ in the synchronous gauge Cartesian coordinates,
\begin{align}\label{Theta_sync}
\Theta( y^\mu) = 3\frac{1}{\bar a^2}\dv{\bar a}{\eta_y} + \frac{1}{2\bar a} \left(\pdv {h_{11}}{\eta_y} +\pdv {h_{22}}{\eta_y} +\pdv {h_{33}}{\eta_y} \right).
\end{align}
From  (\ref{hij}) we deduce the expression for the covariant Hubble monopole in  the synchronous gauge, 
\begin{align}\label{monopolesynch}
   \mathbb{H}^{syn}_0=  \frac{\Theta(x^\mu_s)}{3} &= \frac{1}{\bar a^2}\dv{\bar a}{\eta_y} - \frac{1}{9}\frac{\eta_y}{ \bar a(\eta_y)}\nabla^2 \Phi(y^i).
\end{align}
Because we assumed that the coordinates of linear LTB are exactly the coordinates of the perturbed FLRW in synchronous gauge, this is indeed the monopole of the Hubble parameter in the linear LTB.

We are now in a position to resolve the discrepancy mentioned at the end of \S \ref{linltb} regarding the  covariant Hubble monopole in the  CNG coordinates. Expressing the covariant Hubble monopole in the CNG coordinates, we get
\begin{align}
    \mathbb{H}^N_0 &= \frac{2}{ \eta_y \bar a(\eta_y)} - \frac{1}{9}\frac{ \eta_y}{\bar a(\eta_y)}\nabla^2 \Phi( y^i) \nonumber\\
    &= \frac{2}{\left(\eta + \pdv{L}{\eta}\right)\left(\bar a(\eta) + \frac{2}{\eta}\pdv{L}{\eta}\right)}- \frac{1}{9}\frac{ \eta}{\bar a( \eta)}\nabla^2 \Phi( x^i) \nonumber\\
    &= \frac{2}{\bar a(\eta)\eta}\left(1-\frac{3}{\eta}\pdv{L}{\eta}\right) - \frac{1}{9}\frac{\eta}{\bar a(\eta)}\nabla^2 \Phi \nonumber\\
     &= \frac{2}{\bar a(\eta)\eta}\left(1-\Phi \right) - \frac{1}{9}\frac{\eta}{\bar a(\eta)}\nabla^2 \Phi \nonumber\\
     &= H(1-\Phi) - \frac{1}{9}\frac{\eta}{\bar a}\nabla^2 \Phi. \label{monopolenewton}
\end{align}
The difference between the monopole term of the Hubble in the LLTB (\ref{monopolesynch}) and the Newtonian gauge (\ref{monopolenewton}) is $-H \Phi$. This reflects the fact that the monopole of the Hubble is the only quantity that is directly affected by the gauge transformation of the time coordinate, since its background component is explicitly time dependent. By contrast, the background components of the deceleration ($\Qbb$) and jerk ($\Jbb$) parameters are time independent in the EdS universe. Moreover, their linear-order perturbative corrections are unaffected, and any differences would only appear at second order.

\begin{figure}
    \centering
      \includegraphics[scale=0.45]{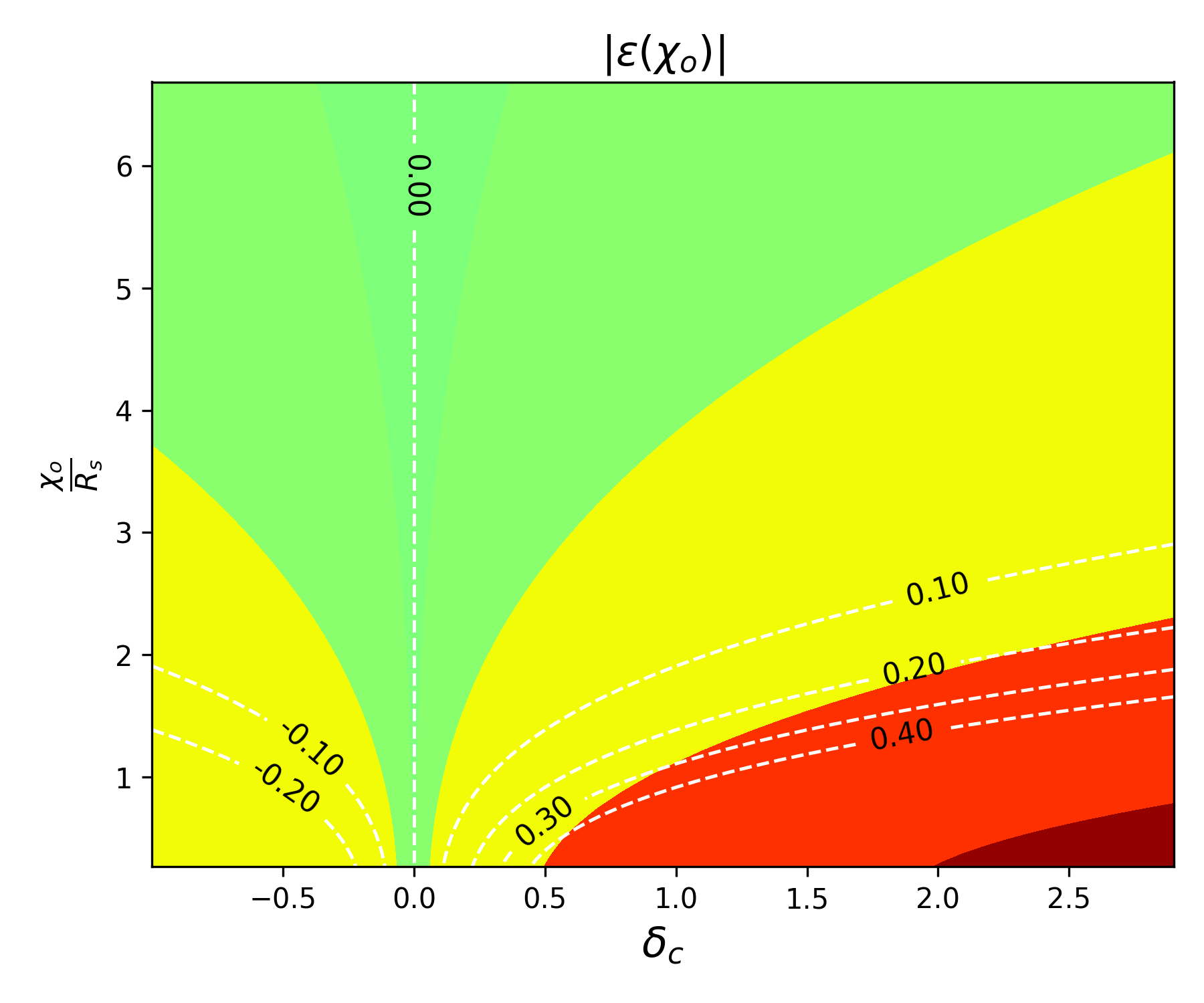}\\
     \hspace{3mm} \includegraphics[scale=0.3]{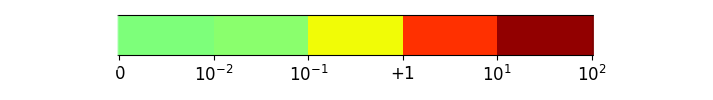}
    \caption{Variation of $\epsilon$ for an off-center observer. Each point in the 2D parameter space corresponds to a specific value of the central density contrast $\delta_c$  and the observer's distance from the center, expressed in units of $ R_s$,  the characteristic size of the inhomogeneity, keeping $R_s$ fixed ($=37.4$ Mpc).}
    \label{multipoles-eta}
\end{figure}

\begin{figure}
  \centering
  \includegraphics[scale=0.41]{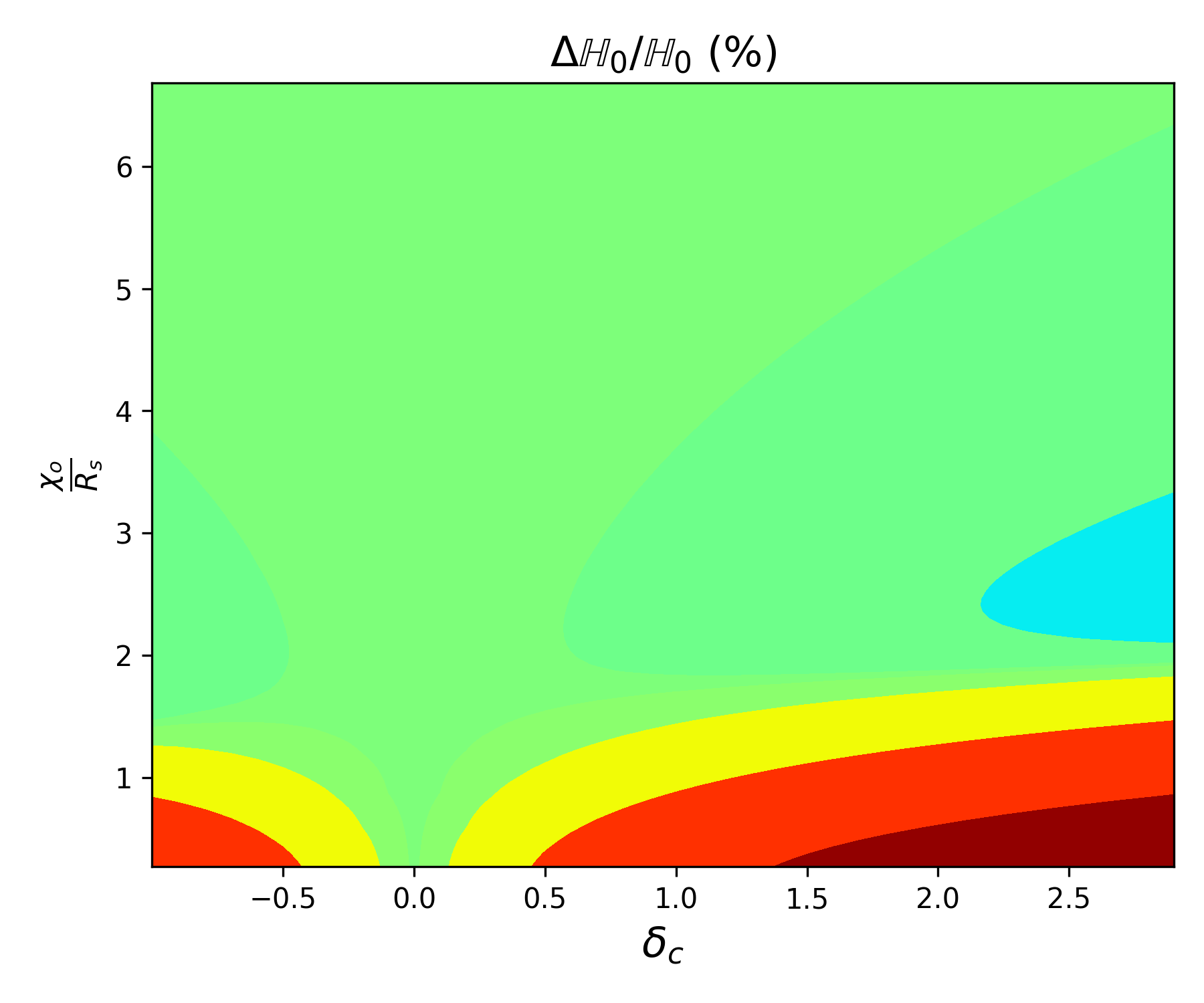}
  \includegraphics[scale=0.41]{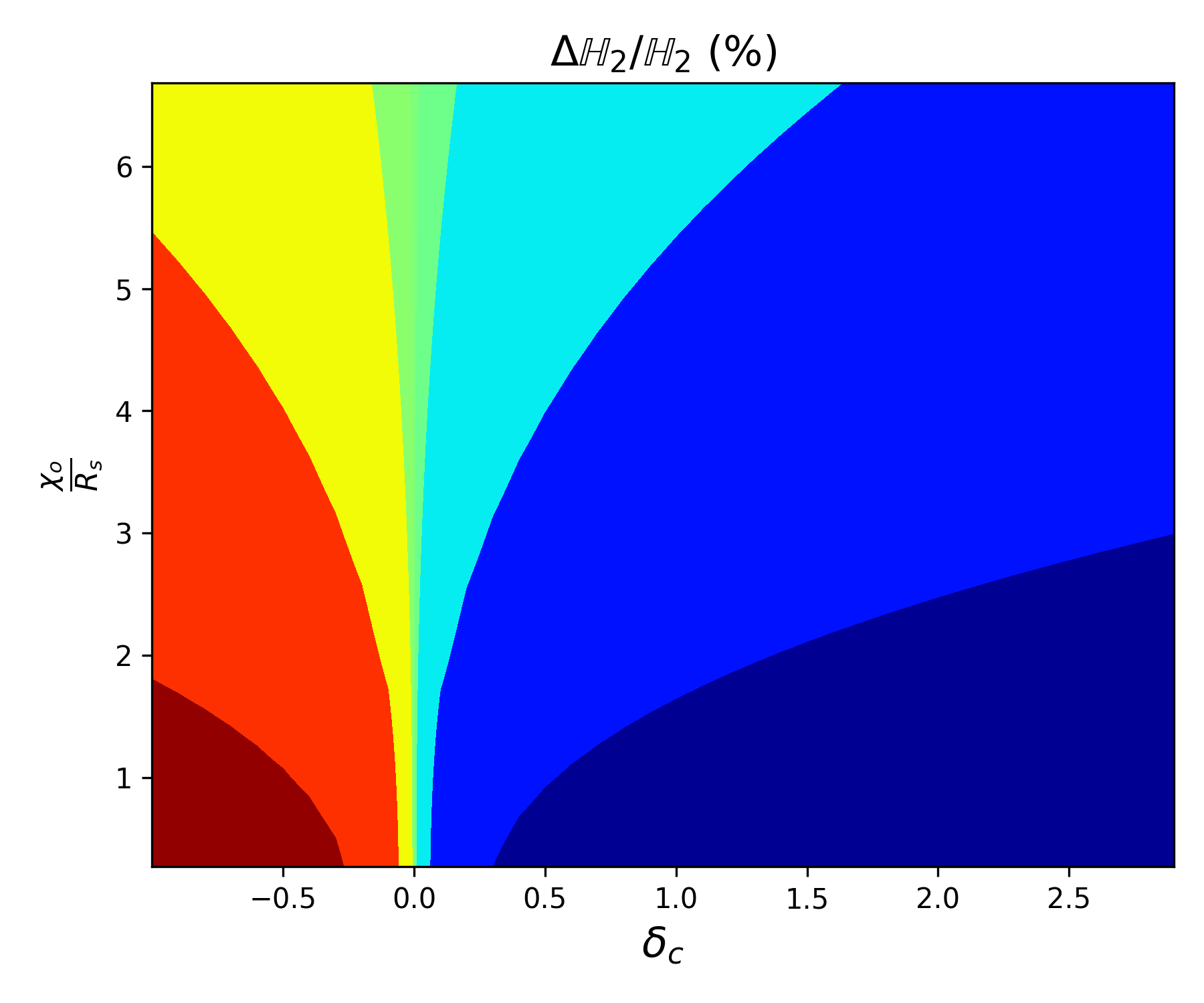}\\
  
  \includegraphics[scale=0.41]{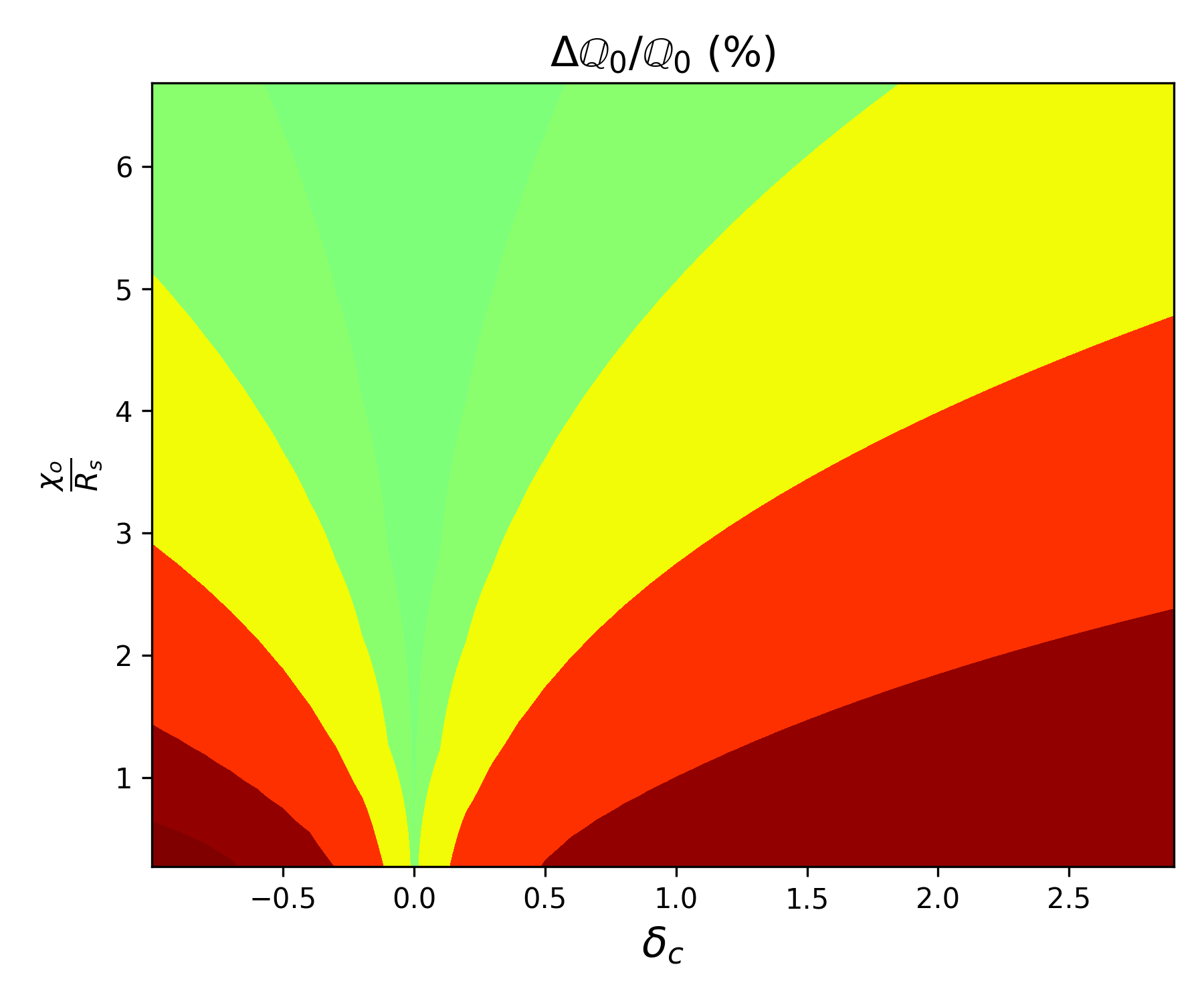}
  \includegraphics[scale=0.41]{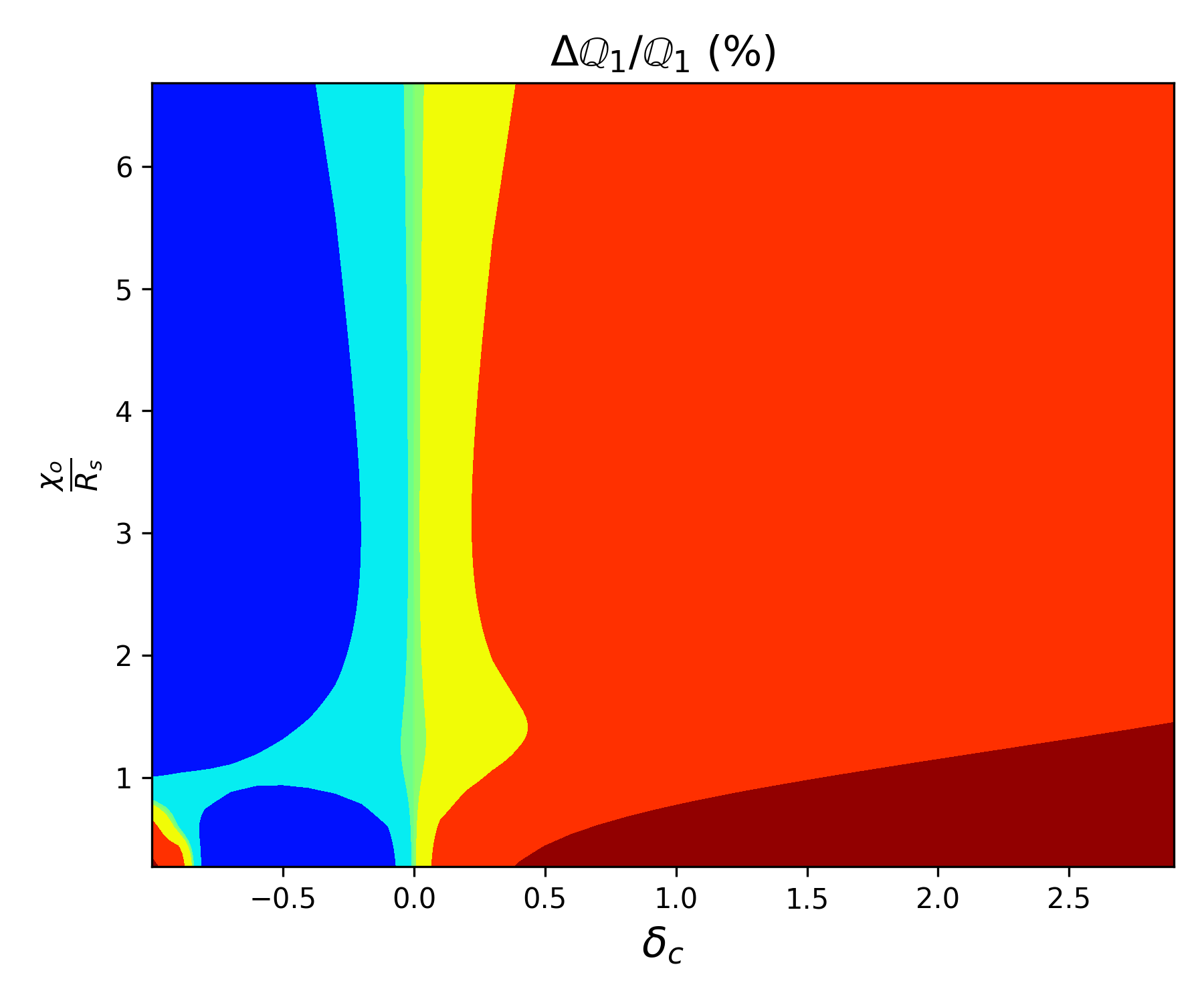}\\

  \includegraphics[scale=0.41]{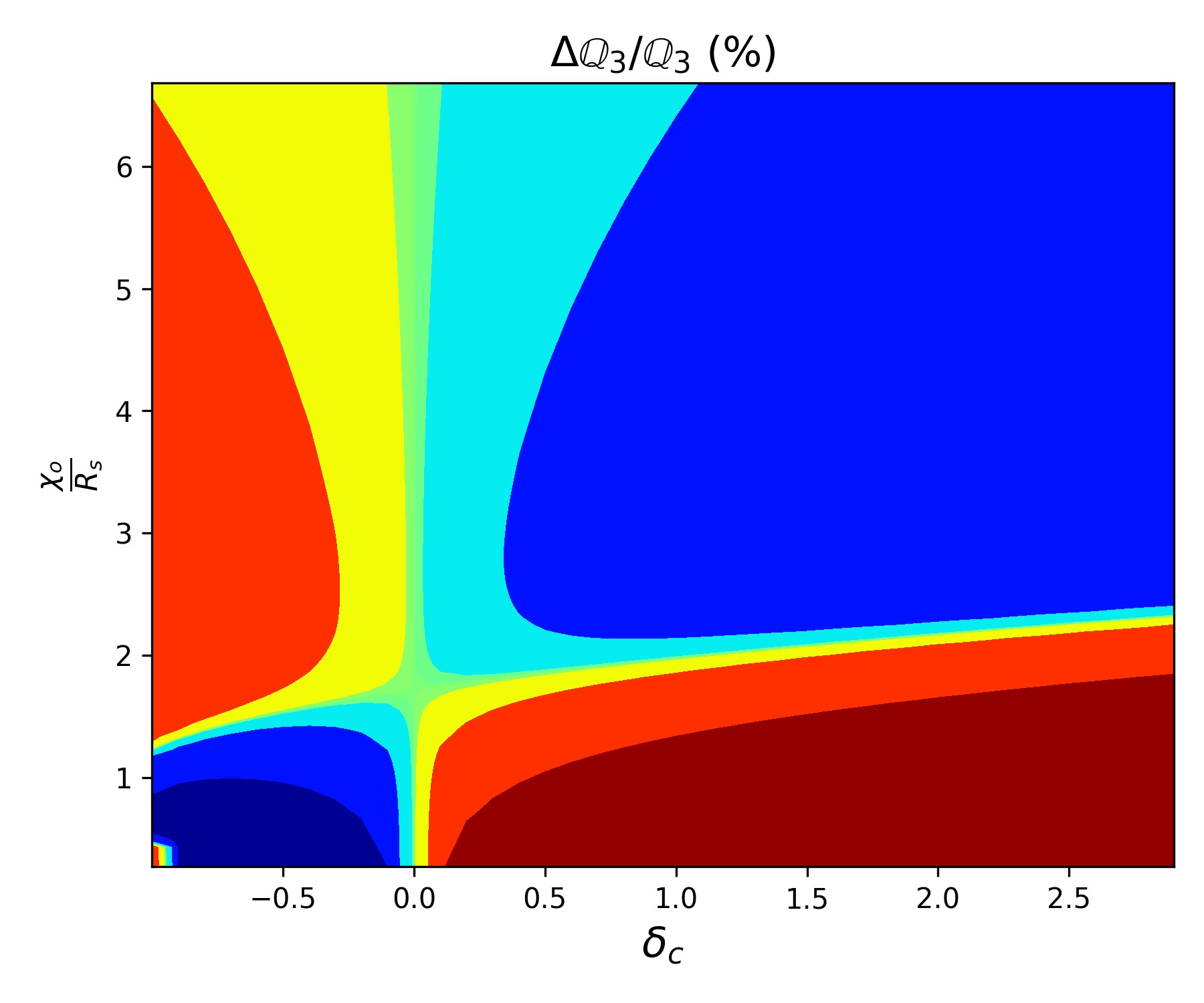}
  \includegraphics[scale=0.41]{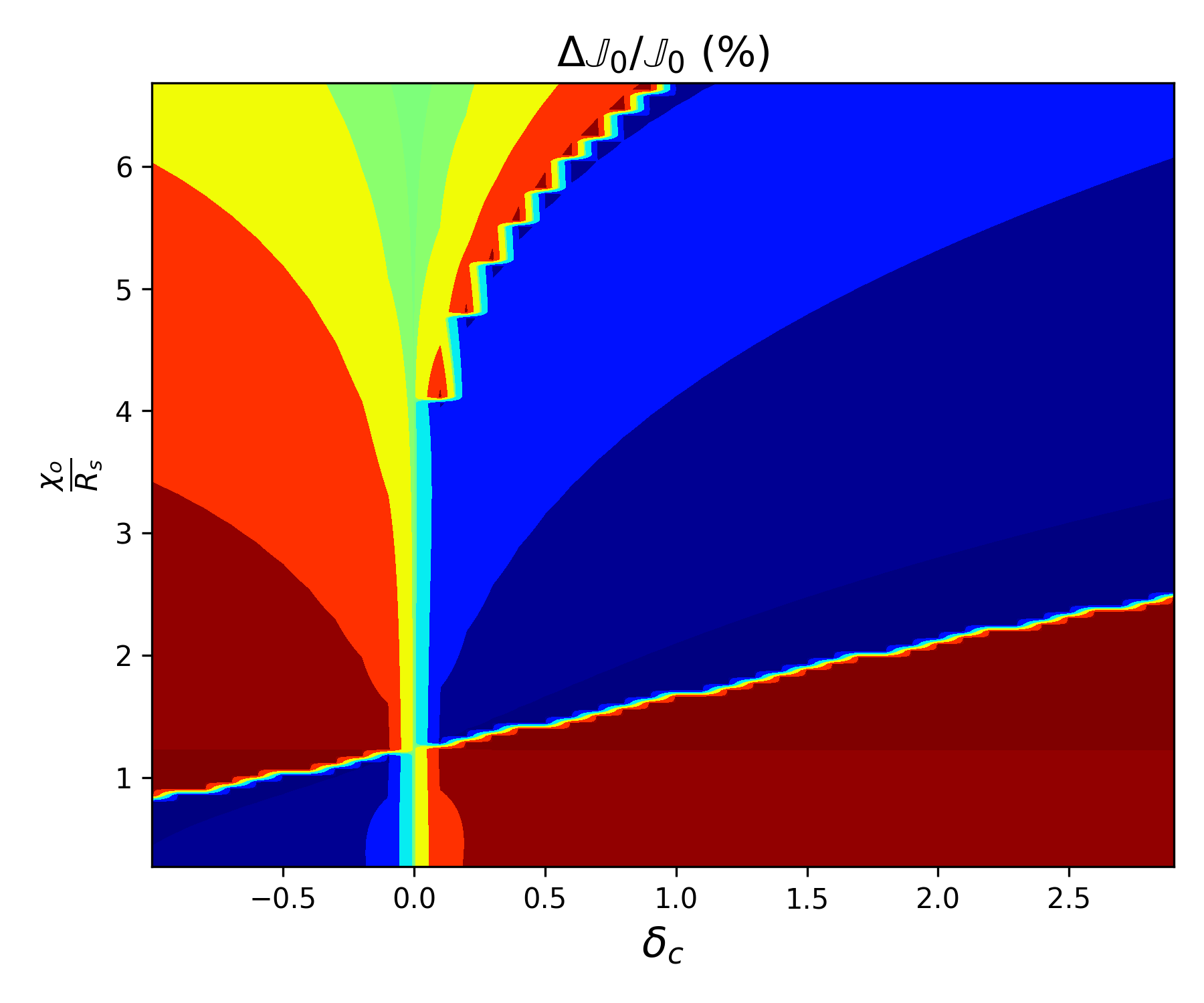}\\
  
  \includegraphics[scale=0.41]{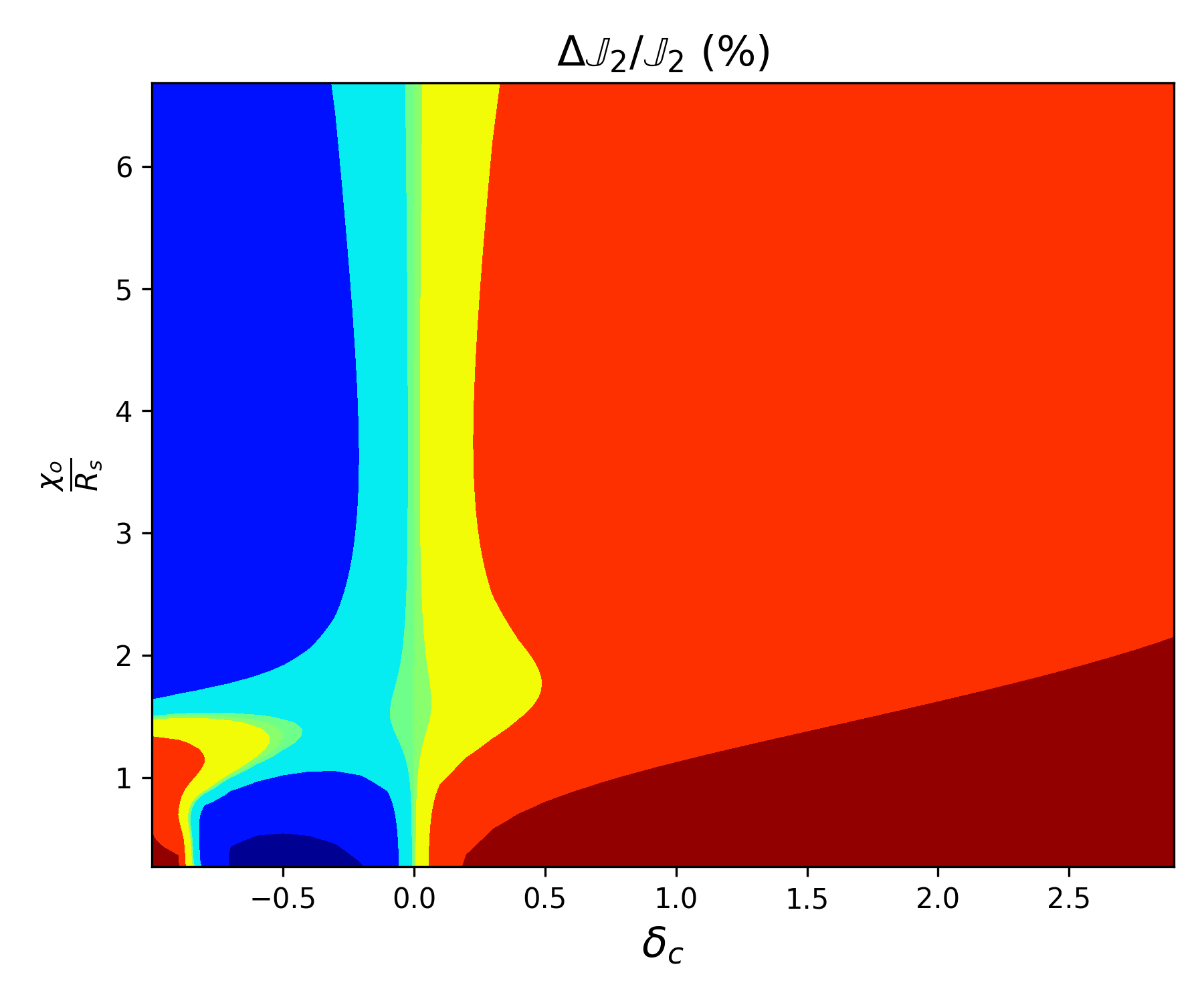}
  \includegraphics[scale=0.41]{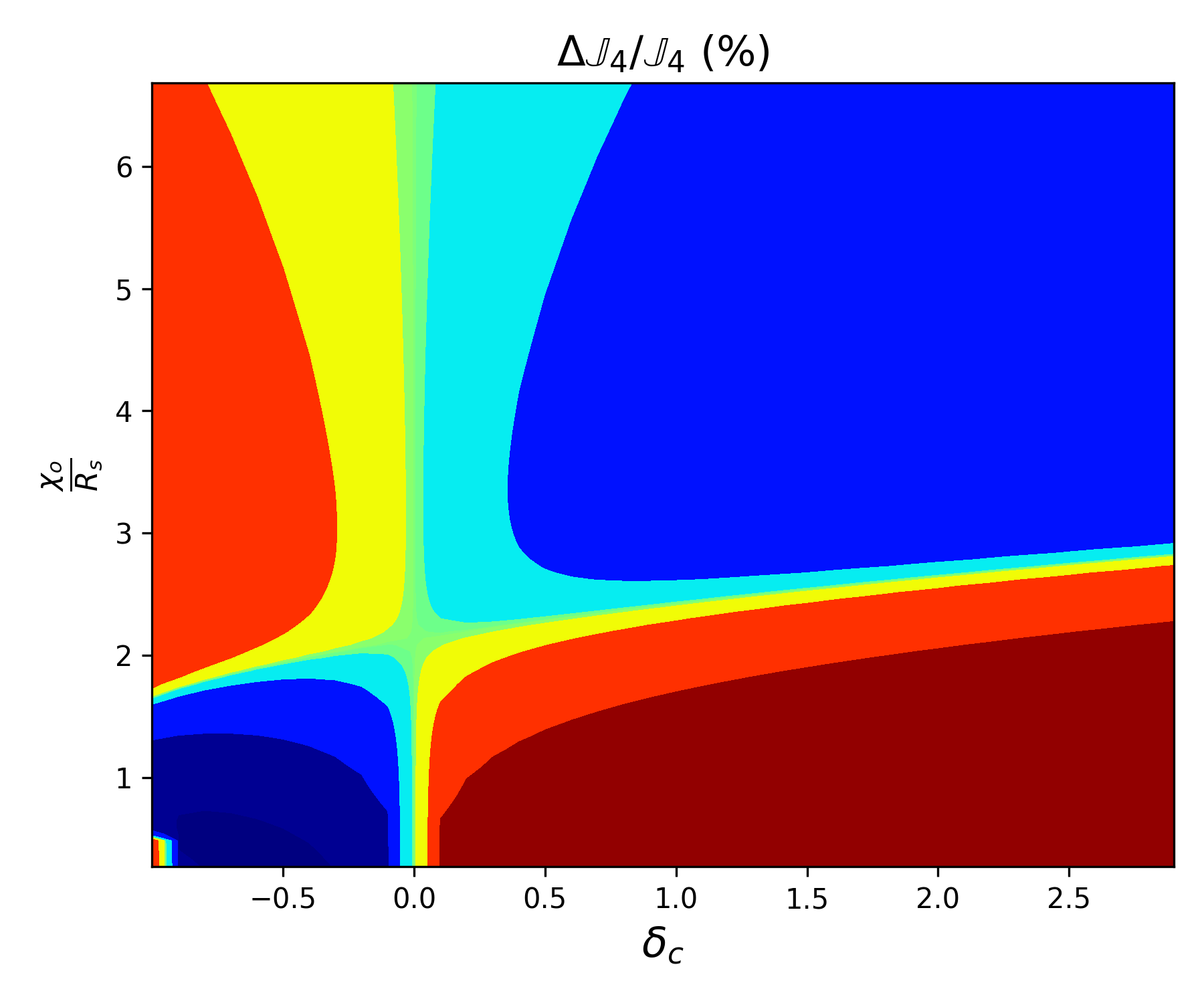}\\  

 \hspace{5mm} \includegraphics[scale=0.41]{fig-6c.png}
   \caption{Deviations of the covariant cosmographic parameters in LTB and LLTB for an off-center observer. Each point in the 2D parameter space corresponds to a specific value of the central density contrast $\delta_c$  and the observer's distance from the center, expressed in units of $ R_s$,  the characteristic size of the inhomogeneity.}
    \label{multipoles}
\end{figure}

In Figure~\ref{multipoles-eta} we examine the behavior of the $\epsilon(\chi_o)$ expansion parameter, which governs the validity of the linear approximation, across various configurations of the observer's position and the density inhomogeneity.
Figure~\ref{multipoles} illustrates how accurately the CC parameters predicted by LPT approximate the true values inferred by an off-center observer in the LTB$_{M1}$ model. As expected, the linear approximation breaks down for high-density peaks when the observer is located near the center of the inhomogeneity. The imprecision introduced in the dominant multipoles of the  CC parameters  due to the gradual breakdown of the linear approximation is shown in the remaining panels.

We find that there is a systematic underestimation, $\mathbb{H}^{lin}_0 < \mathbb{H}_0 $, if the observer is close to the overdensity region $\chi_o  \leq 1.5 R_s$, with the discrepancy  increasing with increasing $|\delta_c|$. On the contrary, as $\chi_o \to \infty$, $\mathbb{H}_0\to\mathbb{H}^{lin}_0$. As for the the quadrupole of the covariant Hubble parameter, we find that the linear quadrupole overestimates the true value,  $\mathbb{H}_2^{lin} > \mathbb{H}_2$, for $\delta_c>0$ and underestimates the true value, $\mathbb{H}_2^{lin} < \mathbb{H}_2$, for all $\delta_c<0$. 

The monopole of the deceleration parameter is systematically underestimated by LPT for all $\delta_c$ and $\chi_o$. Its dipole behaviour is $\mathbb{Q}_1 < \mathbb{Q}_1^{lin}$ for  $\delta_c<0$,    and for $\delta_c>0$, $\mathbb{Q}_1^{lin}<\mathbb{Q}_1$. $\mathbb{Q}_1$ and $\mathbb{Q}_1^{lin}$ cross each other near $\chi_o\approx 0$, when $\delta_c$ varies from $-1 \to -0.5$. This explains the existence of the small linear regime in that region of the contour plot. For the octupole, the existence of the horizontal linear branch in the region $\chi_o \approx R_s \to 2 R_s$ is attributed to crossing of the two curves for a given $\delta_c$. 

 The existence of the unnatural diverging regions in the $\mathbb{J}_0$ contour plot (the jagged lines in Figure \ref{multipoles}), is due to the fact that $\mathbb{J}_0$ and $\mathbb{J}_0^{lin}$ go through $0$, making the ratio numerically unstable. Unlike the differences in $\Hbb_0$ and $\Qbb_0$, the difference in $\Jbb_0$ is dependent on the sign of $\delta_c$. This also explains why the third-order covariant cosmographic expansion is not symmetric in Figure \ref{fig:diag1} (right). $\mathbb{J}_2$ and $\mathbb{J}_2^{lin}$ cross each other near $\chi_o\approx 0$ when $\delta_c$ varies from $-1 \to 0$. This explains the existence of the linear regime in that region of the contour plot. The existence of the linear branch for $\mathbb{J}_4$ is also attributed to crossing of the two curves.

\section{Conclusions}

Motivated by persistent observational tensions—particularly the Hubble constant discrepancy—and by growing evidence of directional anomalies in local expansion-rate measurements, we investigated a non-perturbative, model-independent approach: the covariant cosmography (CC) framework for describing the angular dependence of the distance–redshift relation in the local universe. Our aim was to extend this framework to account for the geometry perceived by an off-center observer in a Lemaître–Tolman–Bondi (LTB) spacetime.

The first objective was to derive analytic expressions for direction-dependent CC quantities—such as the Hubble, deceleration, jerk, and curvature parameters—and to show how their spherical-harmonic components encode observable signatures of anisotropic expansion. The redshift range over which the resulting approximation of the luminosity distance remains valid was tested against the exact solution obtained from the Sachs equation. We found  that in the direction of the structure located around $z\approx 0.045$, the estimation of the cosmographic expansion (up to $\mathcal{O}(z^3)$) is around $5\%$ and diverges rapidly after the structure. In the opposite direction, it remains much more stable and gives an error of $6\%$ at $z\approx0.1$. 

The second objective was to bridge, in the weak-field limit, the non-perturbative covariant cosmography description in LTB spacetimes with the linear perturbation framework in FLRW. We then exploit the resulting dictionary to assess how accurately linear perturbation theory within the standard cosmological model reproduces anisotropies in the distance–redshift relation when the underlying spacetime is instead given by an exact, spherically symmetric LTB metric with an off-center observer. This, in turn, enabled a critical evaluation of the limitations of perturbative techniques in scenarios where the Cosmological Principle may not hold.

This analysis highlights the limitations of perturbative methods when the Cosmological Principle does not strictly hold. For a spherical overdensity, linear perturbation theory already exceeds $10\%$ error in the luminosity distance toward the center for $\delta_c \gtrsim 1$ when the observer lies within the typical size of the structure. In contrast, cosmographic reconstruction maintains better than $10\%$ accuracy up to $\delta_c \lesssim 2.5$. These findings emphasize the need for caution in interpreting cosmological observations under the assumptions of global homogeneity and isotropy.

An important issue deserves discussion. The covariant cosmographic parameters are formally defined as local derivatives of the luminosity distance with respect to redshift at the observer’s position. However, in practice, they must not  be estimated through local differentiation, but via fitting the luminosity distance over a finite redshift range \cite{kalbouneh_marinoni_bel_2023,Kalbouneh:2024,Koksbang_2025,macpherson2025theoreticalpredictiondipolenearby}. This fitting approach treats CC parameters as free parameters optimized to match observational data, thereby reducing sensitivity to local noise and enhancing stability across broader scales. Like the Hubble constant, which is empirically inferred from the slope of the distance–redshift relation rather than from its formal derivative, CC parameters gain physical relevance only through this operational, non-local estimation. This effective approach also allows us to quantify the minimal scale over which data must be coarse-grained in order for the distance–redshift relation to accurately reflect the large-scale geometry of the universe across a significant redshift interval  (see   \cite{paper5} for more details). As a result, the actual imprecision,  which is lower given the average non-local nature of the fitting procedure,  must be estimated by comparing the measured distances with those predicted by CC, using a maximum likelihood analysis.

Looking ahead, we plan to apply the formalism developed in this work to place observational constraints, making use of evidence, such as that presented in \cite{paper5}, on anisotropies in the local expansion-rate field, in order to test their compatibility with an off-center LTB metric and, if so, to extract information about the specific structure of the local spacetime.  On the theoretical side, we plan to extend this approach beyond spherically symmetric models by incorporating more general line elements, thereby introducing additional degrees of freedom for capturing more subtle structures of the local spacetime, as well as for more precise comparison with observational data. Ultimately, this line of research aims to develop a flexible, data-driven cosmological framework that does not rely on the conventional assumptions of the standard model, potentially opening new avenues for addressing the current tensions in our understanding of the universe.

\acknowledgments 

We would like to thank Julien Bel and Federico Piazza 
for useful discussions.  MS, CM and BK are supported by the {\it Agence Nationale de la Recherche} under the grant ANR-24-CE31-6963-01, and the French government under the France 2030 investment plan, as part of the Initiative d’Excellence d'Aix-Marseille Université -  A*MIDEX (AMX-19-IET-012).
RM is supported by the South African Radio Astronomy Observatory and the National Research Foundation (grant no. 75415). 

\newpage

\appendix

\appendix
\section{The optical tidal tensor  for an off-center LTB observer}
\begin{figure}[htbp]
    \centering
    \includegraphics[width=0.75\linewidth]{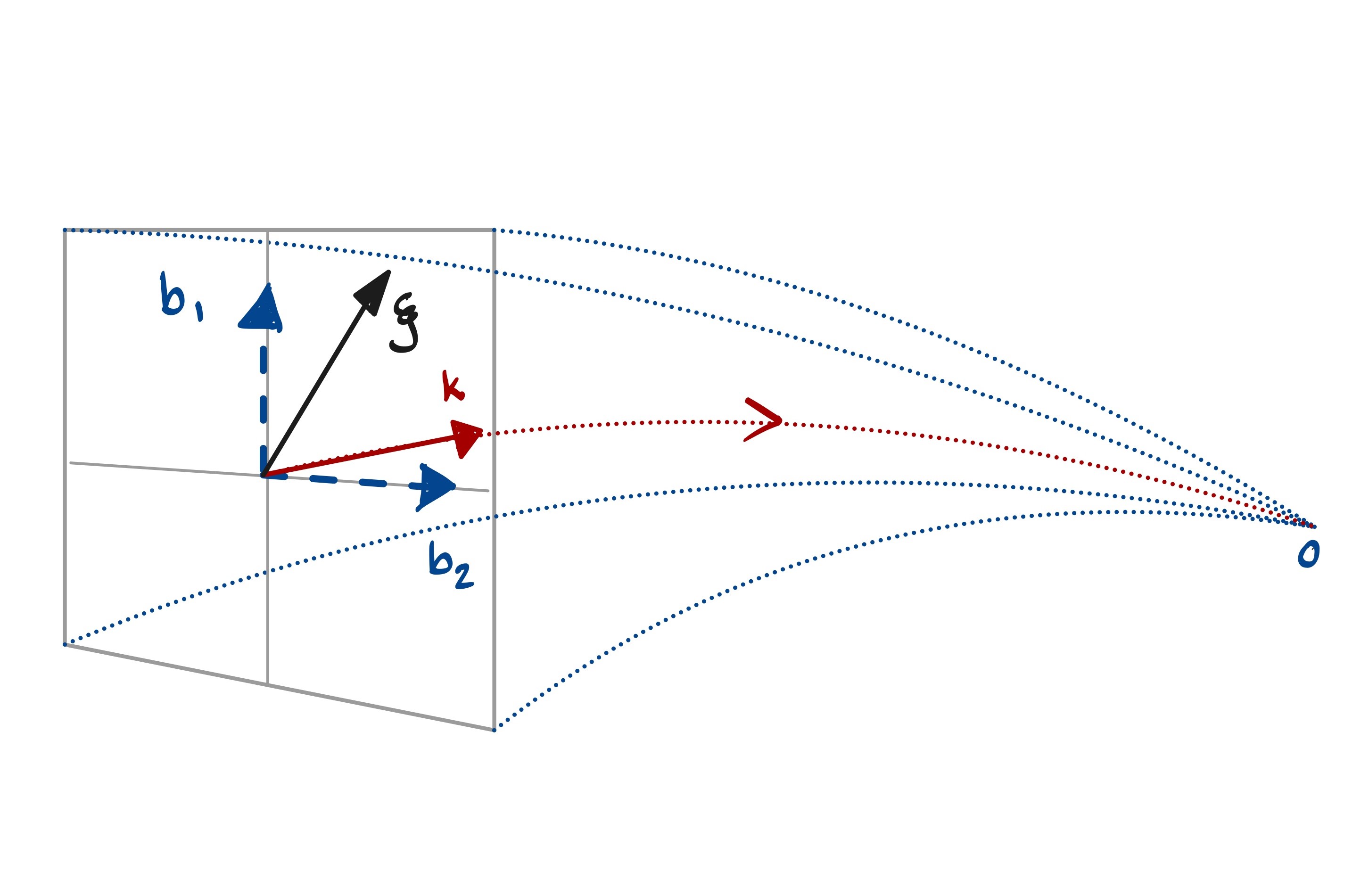}
    \caption{Screen space.}
    \label{fig:enter-label}
\end{figure}

Consider a bundle of null-geodesics emitted by an extended source and converging to the observer at time
$t_0.$
The observer sees the source  subtending    a solid angle  
$\delta \Omega _o$. The cross sectional area of this bundle at the position of the source, $dS_o(t)$,  is equal to the projected surface area of the source at the time 
$t<t_0$ when the photons were emitted.  
The relation  $dS_o(t) \equiv  d_A^2(t) d\Omega_o$
defines  {\it angular diameter distance $ d_A$} of the source  as the {\it area distance}, measured from  the observer  standpoint, and evaluated at  the time  
 $t$ of emission of the photons. 

In order to evaluate  $d_A$ one needs to calculate how the separation vector $\xi$ between two infinitesimally close photons with the same affine parameter $\lambda$, 
 i.e. the transversal size of the null geodesic beam,  changes while propagating in a generic spacetime from the source to the receiver.
To this purpose it is useful to decompose the separation 4-vector $\boldsymbol{\xi}$  into
the Sachs basis $\boldsymbol{b_I}$ (I=1,2) which span the 2D spatial plane  (with metric $\delta_{IJ}$) perpendicular to the 
beam wave-vector $k^{\mu}(\lambda)$ and to the observers line-of-sight  $n^{\mu}$ ( $\boldsymbol{\xi}= \xi^{I} \boldsymbol{b_I}$).
Note that the screen basis vectors $\boldsymbol{b_I}(\lambda)$ are   parallel transported along the beam
$\nabla_{\boldsymbol{k}}(\boldsymbol{b_I}) = 0$
since we assume that the observers are geodesic. 

The components of the separation vector evolves according to the bi-dimensional equivalent of the geodesic deviation equation, i.e. the  Sachs vector equation,
\begin{equation}
 \frac{\dd^2 \xi_{I}}{\dd\lambda^2}=\tau_{IJ}\xi^J ,
 \label{sachsvec15}
 \end{equation}
where 
\begin{equation}
\tau_{IJ} \equiv  R_{\mu \nu \alpha \beta} (b_I)^{\mu}k^{\nu}k^{\alpha}(b_J)^{\beta},
\label{ott}
\end{equation}
is the {\it optical tidal tensor}, a $2 \times 2$ symmetric matrix that connects the evolution of the light bundle with the curvature of spacetime.

A  unique solution of (\ref{sachsvec15}) 
is singled out by  specifying the two initial conditions
\begin{eqnarray}
\xi^{I}_0  & \equiv & \xi^{I}(\lambda=0) ,\nonumber  \\
\dot{\xi}^{I}_0 & \equiv &   \left . \frac{\dd\xi^{I}}{\dd \lambda}\right |_{0}.
\end{eqnarray}
We assume that the  null geodesic bundle  converges at the freely falling terrestrial observer placed at $ \lambda = 0$. This condition  fixes 
$\xi^{I}_0=0$. The  solution can therefore only depend on the initial rate $\dot{\xi}_0^{I}$. 
Since the solution of a linear differential equation depends linearly  on the initial conditions, one can write it as 
\begin{align}
\xi_I(\lambda)= \left . \mathcal{D}_{IJ} \frac{\dd\xi^{J}}{\dd\lambda} \right |_0 .
\end{align}
 One can
 thus recast the problem of determining the evolution of the separation vector $\xi_I$ as the following initial value problem for the unknown  Jacobi map ({\it Sachs equations}):
\begin{eqnarray}
\frac{\dd^2  \mathcal{D}_{IJ}(\lambda) }{\dd \lambda^2} & = & \tau_{IK} \mathcal{D}^{K}_{\phantom{K}J},  \nonumber \\
\left . \mathcal{D}_{IJ}\right|_0 & = & 0,  \nonumber \\
\left . \frac{\dd\mathcal{D}_{IJ}}{\dd\lambda} \right |_0 & = & \delta_{IJ}.
\label{sachs2}
\end{eqnarray}

We can choose the  affine parameter $\lambda$  to coincide with the local Euclidean distance in the observer’s rest frame, so that 
 the initial rate can locally be interpreted as the arrival angle of two rays
\begin{align}  
\theta^I= \left . \frac{\dd\xi^I}{\dd\lambda}\right |_{\lambda =0},
\end{align}
which converge at the position of the observer, i.e. $\theta^I$
 denotes the angle on the celestial sphere between  two rays which were physically separated by $\boldsymbol{\xi}$ at an infinitesimal distance $\dd\lambda$ from the observer.

Integrating (\ref{sachs}) 
 from the observer to a fiducial source located at a position corresponding to the affine parameter 
 $\lambda_s$ leads to
\begin{equation}
\xi_I(\lambda_s)=\left . \mathcal{D}_{IJ}(\lambda_s)\frac{\dd\xi^{J}}{\dd\lambda}\right |_0=\mathcal{D}_{IJ}(\lambda_s)\theta^J .
\label{uiui}
\end{equation}
Therefore the Jacobi matrix represents the transformation between angular coordinates (at the position of the terrestrial observer) into linear metric coordinates (at the source position).
Consequently,  the determinant  $\mathcal{D}$ is  the Jacobian of the coordinate transformation 
$\theta^I \rightarrow \xi^I$, i.e. the ratio between the  infinitesimal `volume' at the source (the physical area of the source $\delta S_o$)  and
the `volume' at the observer position  (the observed solid angle $\delta \Omega_o$). Formally  
\begin{align}
 \left | \mathcal{D}(\lambda_s) \right | =\frac{\delta S_o ( \lambda_s)}{\delta \Omega_o}.   
\end{align}  
This relation can be  turned into an expression that allows us to estimate the  angular diameter distance,
\begin{equation}
d_A(z)= \sqrt{ \left | \mathcal{D}(\lambda_s) \right|}.
\label{jaco}
\end{equation}
The optical tidal tensor  $\tau_{IJ}$ can be decomposed into a pure-trace part and a trace-free part,
\begin{equation}
\tau_{IJ}=-\frac{1}{2}R_{\alpha \beta }k^{\alpha} k^{\beta}\delta_{IJ}+C_{\mu \nu \alpha \beta}(b_I)^{\mu}k^{\nu}k^{\alpha}(b_J)^{\beta} ,
\label{ott2}
\end{equation}
where $R_{\alpha \beta }$ is the Ricci tensor and 
\begin{eqnarray}
C_{\mu \nu \alpha \beta} = R_{\mu \nu \alpha \beta}+\frac{1}{2}\left[g_{\mu \beta} R_{\alpha \nu}- g_{\mu \alpha} R_{\beta \nu}+g_{\nu \alpha}R_{\beta \mu}-g_{\nu \beta}R_{\alpha \mu}\right]
+\frac{R}{6}\left[g_{\mu \alpha}g_{\beta \nu}-g_{\mu \beta}g_{\alpha \nu}\right],
\end{eqnarray}
is the Weyl tensor, 
which is endowed  with the same symmetries as the Riemann tensor $R_{\alpha \beta \mu \nu}$, and is furthermore
 trace-free,  $C^{\alpha}_{\phantom{\alpha} \beta \alpha \gamma}=0.$

Splitting (\ref{ott2}) into the Ricci and Weyl focusing terms 
allows us to better grasp the physical content of the geodesic deviation equation.  
The Ricci focussing originates from matter inside the null bundle and  causes 
$\boldsymbol{\xi}$  to increase or decrease homothetically. 
Note that if gravity is described by the Einstein equations,
\begin{align}
R_{\mu\nu}-\frac{1}{2}g_{\mu \nu}R+\Lambda g_{\mu \nu}=8\pi G  T_{\mu \nu},
\end{align}
then 
\begin{align}
\mathcal{R}=-4\pi G T_{\mu \nu}k^{\mu}k^{\nu}.
\end{align}
In the case of a perfect fluid with rest-frame energy
density $\rho_m$  and pressure $p$, the stress-energy tensor reads 
\begin{equation}
T_{\mu \nu} = (\rho_m  + p)u_{\mu} u_{\nu} + pg_{\mu \nu} ,
\label{pfl}
\end{equation}
so that we get 
\begin{equation}
\mathcal{R} =  - 4 \pi G (\rho_m  + p )(1+z)^2.
\label{riccif}
\end{equation}

On the other hand, the Weyl focussing is a term generated by matter located outside the bundle.  The overall effect of non-local 
contributions is to shear  the beam size. 
The 24 nonzero components  of the  Weyl  tensor
are 
 \begin{eqnarray}
 C_{1212} & = & \frac{C_{1313}}{\sin^2\theta}= \frac{I}{6 \alpha} ,  \nonumber \\
  C_{0202} & = & \frac{C_{0303}}{\sin^2 \theta}=   -\frac{I}{6 \alpha^3},
  \nonumber \\
  C_{2323} & = &- \frac{  A^2  I \sin ^2\theta}{3 \alpha^3} ,     \nonumber \\
  C_{0101} & = &  \frac{I}{3 A^2 \alpha},
 \end{eqnarray}
where 
\begin{align}
I= A \alpha^2 \left(\dot{A} \dot{\alpha}-A \ddot{\alpha}\right)+A A^{\prime} \alpha ^{\prime}-\left(\dot{A}^2-A \ddot{A}+1\right) \alpha^3+\left(A^{\prime 2}-A A^{\prime \prime}\right) \alpha. 
\end{align}

If we assume that the gravitational field satisfies the Einstein field equations with pressureless matter ($p=0$), we obtain
 \begin{eqnarray}
 C_{1212} & = & \frac{C_{1313}}{\sin^2\theta}=\alpha^2   F, \nonumber \\
  C_{0202} & = &  \frac{C_{0303}}{\sin^2 \theta}= -  F ,\nonumber \\
  C_{2323} & = &- 2 A^2 \sin^2 \theta \,  F ,\nonumber \\
  C_{0101} & = &  2\, \frac{\alpha^2 }{A^2}  \,  F ,
 \end{eqnarray}
 where 
\begin{align}
 F =\frac{4}{9}\pi G \, \tilde \rho_m^{\prime}\, \frac{A^3}{A^\prime}.
 \end{align}
Note that the flat average $\tilde \rho_m$ and the  average density $\bar\rho_m$ coincide in LTB models with $k(\chi)=0$.

A normal basis that satisfies the Sachs prescriptions is
\begin{equation}
b_{1}^{\mu}=\beta \left (0,- \frac{A^2}{\alpha^2 }\frac{k^2}{k^1} ,1, 0 \right ), \;\;\;\; \;\;\;\; b_{2}^{\mu}= \gamma \left (0,0,0,1 \right ),
\label{bcent}
\end{equation}
where 
\begin{eqnarray}
\beta & = & \frac{1}{A\sqrt{1+\left( \frac{A k^2}{\alpha k^1}\right)^2}} \;\;\;\;\text{and} \;\;\;\;
\gamma  =  \frac{1}{A \sin \theta}. 
\end{eqnarray}
By using it, we obtain the expression of the Weyl focusing for the off-center observer, 
 \begin{align}
 C_{\mu \nu \alpha \beta}(b_I)^{\mu}k^{\nu}k^{\alpha}(b_J)^{\beta} = \frac{J^2}{A^2}
 \begin{pmatrix}
- F & 0  \\
0  &  F 
\end{pmatrix}.
\end{align}
As a consequence, the optical matrix is diagonal and,  in an LTB cosmology where gravity is sourced by pressureless matter and a cosmological constant,  it is given by  
\begin{equation}\label{tidal tensor2}
\tau^I_{\phantom{I} J}=-4\pi G \rho_m\left[  \left(1+z\right)^2+ \left ( \frac{J}{A} \right )^2 \left(\frac{\rho_m-\tilde \rho_m}{\rho_m} \right)\left(-1\right)^{I-1}\right] \delta^{I}_{\phantom{I}J}.
\end{equation}
We thus conclude that at positions $\chi \sim  \chi_o$, 
the shearing of the beam is a phenomenon of order $(\rho_m-\tilde \rho_m)/\rho_m$. This term is, in general,  non-negligible in inhomogeneous cosmologies, where the local density at position $\chi$ does not coincide with the spherical average value inside  $\chi$.  There might thus be density configurations for which a sizeable shear  imprint is expected. 
There are however important cases where shearing is negligible and the beam expands/contracts isotropically. Indeed, the contribution of the Weyl focusing  vanishes in the spherically symmetric configuration, i.e. when the observer sits at the center of the LTB metric and  thus $J=0$, or if the flat average density $\tilde \rho_m$ equals the local density $\rho_m$ as  in the standard FLRW metric. 

Since the optical matrix  (\ref{tidal tensor2}) is diagonal, the  system (\ref{sachs2})  decouples  and the only non-trivial Sachs equations are 
\begin{eqnarray}
\frac{\dd^2}{\dd \lambda^2}\mathcal{D}_{11}(\lambda) & = & \tau_{11} \mathcal{D}_{11}, \nonumber \\
\frac{\dd^2}{\dd \lambda^2}\mathcal{D}_{22}(\lambda) & = & \tau_{22} \mathcal{D}_{22}, \nonumber \\
\mathcal{D}_{11}(0) & = & \mathcal{D}_{22} = 0, \nonumber \\
\left. \frac{\dd\mathcal{D}_{11}}{\dd\lambda} \right|_0 & = & \left.  \frac{\dd\mathcal{D}_{22}}{\dd\lambda} \right|_0 =1 .
\label{decsac}
\end{eqnarray}
Indeed, since the differential equations are homogeneous,  $\mathcal{D}_{12}(\lambda)=\mathcal{D}_{21}(\lambda)=0$ is the unique solution 
that satisfies  the  given initial conditions.
The angular distance is thus 
 \begin{equation}
d_A(\lambda)=\sqrt{\mathcal{D}_{11}\mathcal{D}_{22}},
 \end{equation}
i.e. the geometric mean of the diagonal terms of the Jacobian matrix $\mathcal{D}_{IJ}$.
As a test, it is straightforward to verify that the angular diameter distance  for a central observer, given by the standard formula $d_A=A(t(\lambda), \chi(\lambda))$,
is indeed the unique solution of (\ref{decsac})  when $J=0$.

\section{Multipoles of $\mathbb{X}^{(4)}$ and $\mathbb{Y}^{(2)}$}\label{appendixsnap}

\begin{align}
       \mathbb{X}^{(4)}_{0}\circeq & \, -2 H_{\|}^2 \dot H_{\|}-\frac{3 \dot H_{\|}''}{5 \alpha ^2}-\frac{3 \alpha '' \dot H_{\|}}{5 \alpha ^3}-\frac{6 \alpha ' \dot H_{\|}'}{5 \alpha
   ^3}-\dot H_{\|}^2+\frac{2 H \ddot {H_{\|}}}{5}+\frac{2 H_{\|} \ddot H }{5}+\frac{104 H \ddot H}{15}+\frac{49 H_{\|}
   \ddot {H_{\|}}}{15}  \nonumber\\
   
   & -\frac{2 \overset{\text{...}}{H}}{3}-\frac{\overset{\text{...}}{H_{\|}}}{3}+\frac{208 H^4}{15}+\frac{8 H^3 H_{\|}}{5}-\frac{56 H^2}{15 \alpha ^2 \chi ^2}+\frac{8 H H''}{5
   \alpha ^2}+\frac{8 H^2 \alpha '}{5 \alpha ^3 \chi }+\frac{16 H^2 H_{\|}^2}{15}+\frac{4 H_{\|} H''}{15 \alpha ^2} \nonumber \\
  
   & -\frac{8 H^2 \dot{H_{\|}}}{15}-\frac{416 \dot{H} H^2}{15}+\frac{32 H H'}{15 \alpha ^2 \chi
   }+\frac{16 \left(H'\right)^2}{15 \alpha ^2}-\frac{8 H \alpha ' H'}{5 \alpha ^3}-\frac{8 H' H_{\|}'}{15 \alpha ^2}-\frac{32 H_{\|} H'}{15 \alpha ^2 \chi }-\frac{4 H_{\|} \alpha ' H'}{15 \alpha ^3} \nonumber\\
   & +\frac{4
   \dot{H}}{5 \alpha ^2 \chi ^2} -\frac{2 \dot{H}''}{5 \alpha ^2}-\frac{4 \dot{H} \alpha '}{5 \alpha ^3 \chi }+\frac{2 \alpha ' \dot{H}'}{5 \alpha ^3}+\frac{8 H H_{\|}^3}{5}-\frac{8 \dot{H} H_{\|}^2}{15}+\frac{16 H
   H_{\|}'}{15 \alpha ^2 \chi }+\frac{56 H H_{\|}}{15 \alpha ^2 \chi ^2}+\frac{8 H H_{\|} \alpha '}{15 \alpha ^3 \chi } \nonumber\\
   & -\frac{12}{5} \dot{H} H H_{\|}-\frac{12}{5} H H_{\|} \dot{H_{\|}}+\frac{4
   \dot{H} \dot{H_{\|}}}{15}+\frac{68 \dot{H}^2}{15}+\frac{88 H_{\|}^4}{15}+\frac{14 H_{\|} H_{\|}''}{5 \alpha ^2}-\frac{32 H_{\|}^2 \alpha '}{15 \alpha ^3 \chi }-\frac{52 H_{\|}^2
   \dot{H_{\|}}}{5} \nonumber\\
   & +\frac{124 H_{\|} H_{\|}'}{15 \alpha ^2 \chi }+\frac{9 \left(H_{\|}'\right)^2}{5 \alpha ^2}-\frac{14 H_{\|} \alpha ' H_{\|}'}{5 \alpha ^3}-\frac{4 \dot{H_{\|}}}{5 \alpha ^2
   \chi ^2}-\frac{2 \dot{H_{\|}}'}{\alpha ^2 \chi }+\frac{3 \dot{H_{\|}} \alpha ''}{5 \alpha ^3}+\frac{4 \dot{H_{\|}} \alpha '}{5 \alpha ^3 \chi }+\frac{9 \alpha ' \dot{H_{\|}}'}{5 \alpha ^3} \nonumber\\
   &+\frac{16
   \dot{H_{\|}}^2}{5}   ,\\

       \mathbb{X}^{(4)}_{1}\circeq & \,      
       -\frac{27 \dot H_{\|} H_{\|}'}{5 \alpha }-\frac{27 H_{\|} \alpha ' \dot H_{\|}}{5 \alpha ^2}-\frac{27 H_{\|} \dot H_{\|}'}{5 \alpha }+\frac{12 \ddot H}{5
   \alpha  \chi }-\frac{6 \ddot H'}{5 \alpha }-\frac{12 \ddot {H_{\|}}}{5 \alpha  \chi }-\frac{9 \ddot {H_{\|}}'}{5 \alpha }-\frac{6 H'''}{35 \alpha ^3}+\frac{464 H^3}{35
   \alpha  \chi } \nonumber \\

   & +\frac{12 H''}{35 \alpha ^3 \chi }+\frac{18 \alpha ' H''}{35 \alpha ^4}-\frac{12 H^2 H_{\|}'}{35 \alpha }+\frac{264 H^2 H_{\|}}{35 \alpha  \chi }+\frac{12 H'}{35 \alpha ^3 \chi ^2}-\frac{18
   \left(\alpha '\right)^2 H'}{35 \alpha ^5}+\frac{6 \alpha '' H'}{35 \alpha ^4}-\frac{8 \alpha ' H'}{7 \alpha ^4 \chi } \nonumber\\

   & +\frac{52 \dot{H} H'}{7 \alpha }+\frac{20 H_{\|}^2 H'}{7 \alpha }-\frac{296 H H_{\|} H'}{35
   \alpha }-\frac{10 \dot{H_{\|}} H'}{7 \alpha }-\frac{88 H^2 H'}{5 \alpha }-\frac{48 H}{35 \alpha ^3 \chi ^3}+\frac{36 H \left(\alpha '\right)^2}{35 \alpha ^5 \chi }-\frac{12 H \alpha ''}{35 \alpha ^4 \chi } \nonumber\\

   & +\frac{44
   H \alpha '}{35 \alpha ^4 \chi ^2}-\frac{568 \dot{H} H}{35 \alpha  \chi }+\frac{276 H \dot{H}'}{35 \alpha }-\frac{296 H H_{\|}^2}{35 \alpha  \chi }-\frac{72 H H_{\|} H_{\|}'}{35 \alpha }+\frac{6 \dot{H}
   H_{\|}'}{35 \alpha }+\frac{148 H \dot{H_{\|}}}{35 \alpha  \chi }-\frac{132 \dot{H} H_{\|}}{35 \alpha  \chi } \nonumber\\

   & +\frac{18 H \dot{H_{\|}}'}{35 \alpha }+\frac{74 H_{\|} \dot{H}'}{35 \alpha }-\frac{3
   H_{\|}'''}{7 \alpha ^3}-\frac{432 H_{\|}^3}{35 \alpha  \chi }-\frac{54 H_{\|}''}{35 \alpha ^3 \chi }+\frac{9 \alpha ' H_{\|}''}{7 \alpha ^4}+\frac{6 H_{\|}'}{7 \alpha ^3 \chi ^2}-\frac{9
   \left(\alpha '\right)^2 H_{\|}'}{7 \alpha ^5} \nonumber\\

   & +\frac{3 \alpha '' H_{\|}'}{7 \alpha ^4}+\frac{96 \alpha ' H_{\|}'}{35 \alpha ^4 \chi }+\frac{498 \dot{H_{\|}} H_{\|}'}{35 \alpha }-\frac{162 H_{\|}^2
   H_{\|}'}{5 \alpha }+\frac{48 H_{\|}}{35 \alpha ^3 \chi ^3}-\frac{36 H_{\|} \left(\alpha '\right)^2}{35 \alpha ^5 \chi }+\frac{12 H_{\|} \alpha ''}{35 \alpha ^4 \chi } \nonumber\\

   & -\frac{44 H_{\|} \alpha '}{35
   \alpha ^4 \chi ^2}+\frac{27 H_{\|} \dot{H_{\|}} \alpha '}{5 \alpha ^2}+\frac{552 H_{\|} \dot{H_{\|}}}{35 \alpha  \chi }+\frac{696 H_{\|} \dot{H_{\|}}'}{35 \alpha } ,\\

    \mathbb{X}^{(4)}_{2}\circeq & \, -4 H_{\|}^2 \dot H_{\|}-\frac{12 \dot H_{\|}''}{7 \alpha ^2}-\frac{12 \alpha '' \dot H_{\|}}{7 \alpha ^3}-\frac{24 \alpha ' \dot H_{\|}'}{7 \alpha ^3}-2
   \dot H_{\|}^2+\frac{2 H \ddot {H_{\|}}}{7}+\frac{2 H_{\|} \ddot H}{7}-\frac{160 H \ddot H}{21}+\frac{148 H_{\|}
   \ddot {H_{\|}}}{21} \nonumber \\
  
   & +\frac{2 \overset{\text{...}}{H}}{3}-\frac{2 \overset{\text{...}}{H_{\|}}}{3}-\frac{368 H^4}{21}+\frac{8 H^3 H_{\|}}{7}+\frac{328 H^2}{21 \alpha ^2 \chi ^2}+\frac{8 H H''}{7
   \alpha ^2}+\frac{8 H^2 \alpha '}{7 \alpha ^3 \chi }+\frac{16 H^2 H_{\|}^2}{21}+\frac{4 H_{\|} H''}{21 \alpha ^2}-\frac{8 H^2 \dot{H_{\|}}}{21} \nonumber\\

   & +\frac{688 \dot{H} H^2}{21}-\frac{352 H H'}{21 \alpha ^2 \chi
   }+\frac{40 \left(H'\right)^2}{21 \alpha ^2}-\frac{8 H \alpha ' H'}{7 \alpha ^3}-\frac{8 H' H_{\|}'}{21 \alpha ^2}+\frac{16 H_{\|} H'}{21 \alpha ^2 \chi }-\frac{4 H_{\|} \alpha ' H'}{21 \alpha ^3}-\frac{44
   \dot{H}}{7 \alpha ^2 \chi ^2}+\frac{24 \dot{H}'}{7 \alpha ^2 \chi } \nonumber\\

   & -\frac{2 \dot{H}''}{7 \alpha ^2}-\frac{4 \dot{H} \alpha '}{7 \alpha ^3 \chi }+\frac{2 \alpha ' \dot{H}'}{7 \alpha ^3}+\frac{8 H H_{\|}^3}{7}-\frac{8
   \dot{H} H_{\|}^2}{21} +\frac{16 H H_{\|}'}{21 \alpha ^2 \chi }-\frac{40 H H_{\|}}{21 \alpha ^2 \chi ^2}+\frac{8 H H_{\|} \alpha '}{21 \alpha ^3 \chi }-\frac{12}{7} \dot{H} H H_{\|} \nonumber\\

   & -\frac{12}{7} H
   H_{\|} \dot{H_{\|}}+\frac{4 \dot{H} \dot{H_{\|}}}{21}-\frac{100 \dot{H}^2}{21}+\frac{304 H_{\|}^4}{21}-\frac{96 H_{\|}^2}{7 \alpha ^2 \chi ^2}+\frac{8 H_{\|} H_{\|}''}{\alpha ^2}-\frac{32
   H_{\|}^2 \alpha '}{21 \alpha ^3 \chi }-\frac{172 H_{\|}^2 \dot{H_{\|}}}{7} \nonumber\\

   & +\frac{124 H_{\|} H_{\|}'}{21 \alpha ^2 \chi }+\frac{36 \left(H_{\|}'\right)^2}{7 \alpha ^2}-\frac{8 H_{\|} \alpha '
   H_{\|}'}{\alpha ^3}+\frac{44 \dot{H_{\|}}}{7 \alpha ^2 \chi ^2}-\frac{10 \dot{H_{\|}}'}{7 \alpha ^2 \chi }+\frac{12 \dot{H_{\|}} \alpha ''}{7 \alpha ^3}+\frac{4 \dot{H_{\|}} \alpha '}{7 \alpha ^3
   \chi }+\frac{36 \alpha ' \dot{H_{\|}}'}{7 \alpha ^3} \nonumber\\

   & +\frac{46 \dot{H_{\|}}^2}{7},\\

       \mathbb{X}^{(4)}_{3}\circeq & \,  -\frac{18 \dot H_{\|} H_{\|}'}{5 \alpha }-\frac{18 H_{\|} \alpha ' \dot H_{\|}}{5 \alpha ^2}-\frac{18 H_{\|} \dot H_{\|}'}{5 \alpha }-\frac{12 \ddot H}{5
   \alpha  \chi }+\frac{6 \ddot H'}{5 \alpha }+\frac{12 \ddot {H_{\|}}}{5 \alpha  \chi }-\frac{6 \ddot {H_{\|}}'}{5 \alpha }+\frac{2 H'''}{45 \alpha ^3}-\frac{688 H^3}{45
   \alpha  \chi } \nonumber\\

   & +\frac{4 H''}{5 \alpha ^3 \chi }-\frac{2 \alpha ' H''}{15 \alpha ^4}+\frac{4 H^2 H_{\|}'}{45 \alpha }-\frac{248 H^2 H_{\|}}{45 \alpha  \chi }-\frac{68 H'}{15 \alpha ^3 \chi ^2}+\frac{2 \left(\alpha
   '\right)^2 H'}{15 \alpha ^5}-\frac{2 \alpha '' H'}{45 \alpha ^4}-\frac{8 \alpha ' H'}{9 \alpha ^4 \chi }-\frac{68 \dot{H} H'}{9 \alpha } \nonumber\\

   & -\frac{28 H_{\|}^2 H'}{9 \alpha }+\frac{104 H H_{\|} H'}{15 \alpha
   }+\frac{14 \dot{H_{\|}} H'}{9 \alpha }+\frac{872 H^2 H'}{45 \alpha }+\frac{112 H}{15 \alpha ^3 \chi ^3}-\frac{4 H \left(\alpha '\right)^2}{15 \alpha ^5 \chi }+\frac{4 H \alpha ''}{45 \alpha ^4 \chi }+\frac{92 H
   \alpha '}{45 \alpha ^4 \chi ^2} \nonumber\\

   & +\frac{776 \dot{H} H}{45 \alpha  \chi }-\frac{124 H \dot{H}'}{15 \alpha }+\frac{472 H H_{\|}^2}{45 \alpha  \chi }+\frac{8 H H_{\|} H_{\|}'}{15 \alpha }-\frac{2 \dot{H}
   H_{\|}'}{45 \alpha }-\frac{236 H \dot{H_{\|}}}{45 \alpha  \chi }+\frac{124 \dot{H} H_{\|}}{45 \alpha  \chi }-\frac{2 H \dot{H_{\|}}'}{15 \alpha } \nonumber\\

   & -\frac{26 H_{\|} \dot{H}'}{15 \alpha }-\frac{4
   H_{\|}'''}{9 \alpha ^3}+\frac{464 H_{\|}^3}{45 \alpha  \chi }+\frac{2 H_{\|}''}{5 \alpha ^3 \chi }+\frac{4 \alpha ' H_{\|}''}{3 \alpha ^4}+\frac{10 H_{\|}'}{3 \alpha ^3 \chi ^2}-\frac{4
   \left(\alpha '\right)^2 H_{\|}'}{3 \alpha ^5}+\frac{4 \alpha '' H_{\|}'}{9 \alpha ^4}-\frac{32 \alpha ' H_{\|}'}{45 \alpha ^4 \chi } \nonumber\\

   & +\frac{434 \dot{H_{\|}} H_{\|}'}{45 \alpha }-\frac{1072
   H_{\|}^2 H_{\|}'}{45 \alpha }-\frac{112 H_{\|}}{15 \alpha ^3 \chi ^3}+\frac{4 H_{\|} \left(\alpha '\right)^2}{15 \alpha ^5 \chi }-\frac{4 H_{\|} \alpha ''}{45 \alpha ^4 \chi }-\frac{92 H_{\|}
   \alpha '}{45 \alpha ^4 \chi ^2}+\frac{18 H_{\|} \dot{H_{\|}} \alpha '}{5 \alpha ^2}  \nonumber\\
       
    & -\frac{664 H_{\|} \dot{H_{\|}}}{45 \alpha  \chi }+\frac{206 H_{\|} \dot{H_{\|}}'}{15 \alpha }
       ,\\

       \mathbb{X}^{(4)}_{4}\circeq & \, -\frac{24 \dot H_{\|}''}{35 \alpha ^2}-\frac{24 \alpha '' \dot H_{\|}}{35 \alpha ^3}-\frac{48 \alpha ' \dot H_{\|}'}{35 \alpha ^3}-\frac{24 H \ddot {H_{\|}}}{35}-\frac{24 H_{\|} \ddot H}{35}+\frac{24 H
   \ddot H}{35}+\frac{24 H_{\|} \ddot {H_{\|}}}{35}+\frac{128 H^4}{35}-\frac{96 H^3 H_{\|}}{35} \nonumber\\
   
   & -\frac{416 H^2}{35 \alpha ^2 \chi ^2}-\frac{96 H H''}{35 \alpha ^2}-\frac{96 H^2 \alpha '}{35 \alpha ^3 \chi }-\frac{64 H^2
   H_{\|}^2}{35}-\frac{16 H_{\|} H''}{35 \alpha ^2}+\frac{32 H^2 \dot{H_{\|}}}{35}-\frac{176 \dot{H} H^2}{35}+\frac{512 H H'}{35 \alpha ^2 \chi } \nonumber\\

   & -\frac{104 \left(H'\right)^2}{35 \alpha ^2}+\frac{96 H \alpha ' H'}{35 \alpha ^3}+\frac{32 H' H_{\|}'}{35 \alpha
   ^2}+\frac{48 H_{\|} H'}{35 \alpha ^2 \chi }+\frac{16 H_{\|} \alpha ' H'}{35 \alpha ^3}+\frac{192 \dot{H}}{35 \alpha ^2 \chi ^2}-\frac{24 \dot{H}'}{7 \alpha ^2 \chi }+\frac{24 \dot{H}''}{35 \alpha ^2} \nonumber\\

   & +\frac{48 \dot{H} \alpha '}{35 \alpha ^3 \chi }-\frac{24 \alpha '
   \dot{H}'}{35 \alpha ^3}-\frac{96 H H_{\|}^3}{35}+\frac{32 \dot{H} H_{\|}^2}{35}-\frac{64 H H_{\|}'}{35 \alpha ^2 \chi }-\frac{64 H H_{\|}}{35 \alpha ^2 \chi ^2}-\frac{32 H H_{\|} \alpha '}{35 \alpha ^3 \chi }+\frac{144}{35} \dot{H} H
   H_{\|} \nonumber\\

   & +\frac{144}{35} H H_{\|} \dot{H_{\|}}-\frac{16 \dot{H} \dot{H_{\|}}}{35}+\frac{8 \dot{H}^2}{35}+\frac{128 H_{\|}^4}{35}+\frac{96 H_{\|}^2}{7 \alpha ^2 \chi ^2}+\frac{16 H_{\|} H_{\|}''}{5 \alpha ^2}+\frac{128 H_{\|}^2 \alpha '}{35
   \alpha ^3 \chi }-\frac{176 H_{\|}^2 \dot{H_{\|}}}{35} \nonumber\\

   & -\frac{496 H_{\|} H_{\|}'}{35 \alpha ^2 \chi }+\frac{72 \left(H_{\|}'\right)^2}{35 \alpha ^2}-\frac{16 H_{\|} \alpha ' H_{\|}'}{5 \alpha ^3}-\frac{192 \dot{H_{\|}}}{35 \alpha ^2 \chi
   ^2}+\frac{24 \dot{H_{\|}}'}{7 \alpha ^2 \chi }+\frac{24 \dot{H_{\|}} \alpha ''}{35 \alpha ^3}-\frac{48 \dot{H_{\|}} \alpha '}{35 \alpha ^3 \chi }+\frac{72 \alpha ' \dot{H_{\|}}'}{35 \alpha ^3} \nonumber\\

   & +\frac{8 \dot{H_{\|}}^2}{35},\\

       \mathbb{X}^{(4)}_{5}\circeq & \, \frac{8 H'''}{63 \alpha ^3}+\frac{128 H^3}{63 \alpha  \chi }-\frac{8 H''}{7 \alpha ^3 \chi }-\frac{8 \alpha ' H''}{21 \alpha ^4}+\frac{16 H^2 H_{\|}'}{63 \alpha }-\frac{128 H^2 H_{\|}}{63 \alpha  \chi }+\frac{88 H'}{21 \alpha ^3 \chi ^2}+\frac{8 \left(\alpha
   '\right)^2 H'}{21 \alpha ^5}-\frac{8 \alpha '' H'}{63 \alpha ^4} \nonumber\\
   
   & +\frac{128 \alpha ' H'}{63 \alpha ^4 \chi }+\frac{8 \dot{H} H'}{63 \alpha }+\frac{16 H_{\|}^2 H'}{63 \alpha }+\frac{32 H H_{\|} H'}{21 \alpha }-\frac{8 \dot{H_{\|}} H'}{63 \alpha }-\frac{16 H^2 H'}{9
   \alpha }-\frac{128 H}{21 \alpha ^3 \chi ^3}-\frac{16 H \left(\alpha '\right)^2}{21 \alpha ^5 \chi } \nonumber\\

   & +\frac{16 H \alpha ''}{63 \alpha ^4 \chi }-\frac{208 H \alpha '}{63 \alpha ^4 \chi ^2}-\frac{64 \dot{H} H}{63 \alpha  \chi }+\frac{8 H \dot{H}'}{21 \alpha }-\frac{128 H
   H_{\|}^2}{63 \alpha  \chi }+\frac{32 H H_{\|} H_{\|}'}{21 \alpha }-\frac{8 \dot{H} H_{\|}'}{63 \alpha }+\frac{64 H \dot{H_{\|}}}{63 \alpha  \chi } \nonumber\\
   
   & +\frac{64 \dot{H} H_{\|}}{63 \alpha  \chi }-\frac{8 H \dot{H_{\|}}'}{21 \alpha }-\frac{8 H_{\|}
   \dot{H}'}{21 \alpha }-\frac{8 H_{\|}'''}{63 \alpha ^3}+\frac{128 H_{\|}^3}{63 \alpha  \chi }+\frac{8 H_{\|}''}{7 \alpha ^3 \chi }+\frac{8 \alpha ' H_{\|}''}{21 \alpha ^4}-\frac{88 H_{\|}'}{21 \alpha ^3 \chi ^2} -\frac{8 \left(\alpha '\right)^2
   H_{\|}'}{21 \alpha ^5}  \nonumber\\

   & +\frac{8 \alpha '' H_{\|}'}{63 \alpha ^4}-\frac{128 \alpha ' H_{\|}'}{63 \alpha ^4 \chi }+\frac{8 \dot{H_{\|}} H_{\|}'}{63 \alpha }-\frac{16 H_{\|}^2 H_{\|}'}{9 \alpha }+\frac{128 H_{\|}}{21 \alpha ^3 \chi ^3}+\frac{16
   H_{\|} \left(\alpha '\right)^2}{21 \alpha ^5 \chi }-\frac{16 H_{\|} \alpha ''}{63 \alpha ^4 \chi }+\frac{208 H_{\|} \alpha '}{63 \alpha ^4 \chi ^2} \nonumber\\

   &-\frac{64 H_{\|} \dot{H_{\|}}}{63 \alpha  \chi }+\frac{8 H_{\|} \dot{H_{\|}}'}{21 \alpha }   ,\\

\mathbb{Y}^{(2)}_{0} \circeq & \, \frac{4 \ddot{H}}{3}+\frac{2 \ddot{H_{\|}}}{3}+\frac{8 H''}{3 \alpha ^2}+\frac{8 H'}{\alpha ^2 \chi }-\frac{8 \alpha ' H'}{3 \alpha ^3}-\frac{8 H \dot{H_{\|}}}{3}-\frac{8 \dot{H} H_{\|}}{3}+\frac{8 H}{3 \chi ^2}-\frac{8 H_{\|}'}{3 \alpha ^2 \chi }-\frac{8 H_{\|}}{3 \alpha ^2 \chi ^2}+\frac{16 H_{\|} \alpha
   '}{3 \alpha ^3 \chi } \nonumber\\
    
    & +\frac{4 H_{\|} \dot{H_{\|}}}{3} ,\\

\mathbb{Y}^{(2)}_{1} \circeq & \, -\frac{8}{5 \alpha ^3 \chi ^3}+\frac{24 \left(\alpha '\right)^2}{5 \alpha ^5 \chi }-\frac{8 \alpha ''}{5 \alpha ^4 \chi }+\frac{8}{5 \alpha  \chi ^3}+\frac{24 H H'}{5 \alpha }-\frac{16 H_{\|} H'}{\alpha }+\frac{24 \dot{H}}{5 \alpha  \chi }+\frac{28 \dot{H}'}{5 \alpha }-\frac{8 H H_{\|}'}{5 \alpha } \nonumber\\

& -\frac{48 H H_{\|}}{5 \alpha  \chi } +\frac{48
   H_{\|}^2}{5 \alpha  \chi }+\frac{4 H_{\|} H_{\|}'}{5 \alpha }-\frac{24 \dot{H_{\|}}}{5 \alpha  \chi }+\frac{2 \dot{H_{\|}}'}{5 \alpha },\\

\mathbb{Y}^{(2)}_{2} \circeq & \, \frac{2 \ddot{H}}{3}-\frac{2 \ddot{H_{\|}}}{3}+\frac{10 H''}{3 \alpha ^2}-4 H^2 H_{\|}-\frac{10 \alpha ' H'}{3 \alpha ^3}-\frac{4 H}{\alpha ^2 \chi ^2}-\frac{4 H \alpha '}{\alpha ^3 \chi }+4 H H_{\|}^2+\frac{2 H \dot{H_{\|}}}{3}-\frac{10 \dot{H} H_{\|}}{3} \nonumber\\

& -\frac{8 H}{3 \chi ^2}+4 \dot{H} H-\frac{10
   H_{\|}'}{3 \alpha ^2 \chi }+\frac{20 H_{\|}}{3 \alpha ^2 \chi ^2}+\frac{20 H_{\|} \alpha '}{3 \alpha ^3 \chi }-\frac{4 H_{\|} \dot{H_{\|}}}{3} ,\\

\mathbb{Y}^{(2)}_{3} \circeq & \, -\frac{8}{5 \alpha ^3 \chi ^3}-\frac{6 \left(\alpha '\right)^2}{5 \alpha ^5 \chi }+\frac{2 \alpha ''}{5 \alpha ^4 \chi }-\frac{2 \alpha '}{\alpha ^4 \chi ^2}+\frac{8}{5 \alpha  \chi ^3}-\frac{16 H H'}{5 \alpha } +\frac{2 H_{\|} H'}{\alpha }+\frac{4 \dot{H}}{5 \alpha  \chi }-\frac{2 \dot{H}'}{5 \alpha }+\frac{2 H H_{\|}'}{5 \alpha } \nonumber\\

& -\frac{12 H_{\|}^2}{5 \alpha  \chi }+\frac{4 H_{\|} H_{\|}'}{5 \alpha }-\frac{4 \dot{H_{\|}}}{5 \alpha  \chi }+\frac{2 \dot{H_{\|}}'}{5 \alpha }.
\end{align}

For the case of the central observer, we have,
\begin{align}
    \mathbb{X}^{(4)}_{0} \circeq & \, -6 \ddot H'+11 H \ddot H-\dddot H -2 H^2 H' \Omega _{k}+24 H^4-4H'''+42 H H''-46 \dot H H^2+36 H'^2+ 30 \dot H H' \nonumber\\
   & -116 H^2 H'+50 H \dot H'-9 \dot H''+7 \dot H^2, \\
   \mathbb{Y}^{(2)}_{0} \circeq & \, 2\ddot H +2\dot H' -4H\dot H -2 HH' -8H^3\Omega_k + 6HH'\Omega_k +3H^2\Omega_k'.
\end{align}

\bibliographystyle{JHEP}
\bibliography{main}
\end{document}